\documentclass[12pt]{article}

\usepackage{amssymb,latexsym,amsmath}

\usepackage{fullpage}
\usepackage{nicefrac}

\usepackage{graphicx}

\makeatletter \@addtoreset{equation}{section}
\makeatother

\newcommand{\sh}{{\rm sh}}
\newcommand{\ch}{{\rm ch}}
\newcommand{\shsh}{{\rm sh}^2}
\newcommand{\chch}{{\rm ch}^2}
\newcommand{\txt}[1]{{\textstyle{ #1}}}

\begin{document}
\begin{titlepage}
	\thispagestyle{empty}
	\begin{flushright}
		\hfill{DFPD-2010/TH/14}
	\end{flushright}
	
	\vspace{55pt}  
	 
	\begin{center}
	    { \Huge{A supersymmetric consistent truncation\\ [2mm] for conifold solutions}}
		
		\vspace{30pt}
		
		{Davide Cassani$^{\,1}$ and Anton F. Faedo$^{\,2}$}
		
		\vspace{25pt}
		
		{\small
		{\it ${}^1$  Dipartimento di Fisica ``Galileo Galilei''\\
		Universit\`a di Padova, Via Marzolo 8, 35131 Padova, Italy}
		
		\vspace{15pt}
		
		{\it ${}^2$  INFN, Sezione di Padova \\
		Via Marzolo 8, 35131 Padova, Italy}
		
		\vspace{15pt}

		{\tt cassani, faedo AT pd.infn.it}
		}
		
		\vspace{70pt}

\begin{abstract}	

We establish a supersymmetric consistent truncation of type IIB supergravity on the 
$T^{1,1}$ coset space, based on extending the Papadopoulos--Tseytlin ansatz to the full set of SU(2)$\times$SU(2) invariant Kaluza--Klein modes. The five-dimensional model is a gauged $\mathcal N=4$ supergravity with three vector multiplets, which incorporates various conifold solutions and is suitable for the study of their dynamics.
By analysing the scalar potential we find a family of new non-supersymmetric AdS$_5$ extrema interpolating between a solution obtained long ago by Romans and a solution employing an Einstein metric on $T^{1,1}$ different from the standard one. Finally, we discuss some simple consistent subtruncations preserving $\mathcal N=2$ supersymmetry. One of them still \hbox{contains} the Klebanov--Strassler solution, and is compatible with the inclusion of smeared \hbox{D7-branes}.

\end{abstract}
	
\end{center}
	
\vspace{10pt}

\end{titlepage}

\baselineskip 5.5 mm


\section{Introduction}

Starting with the work of \cite{KlebanovWitten}, the conifold \cite {CandelasOssaConifolds} has played a prominent role in the study of gauge/gravity dualities preserving $\mathcal N=1$ supersymmetry on the field theory side. This manifold is a simple non-compact Calabi--Yau 3-fold, which can be seen as a cone over the homogeneous space $T^{1,1}=\frac{{\rm SU}(2)\times {\rm SU}(2)}{{\rm U}(1)}$ endowed with its Sasaki--Einstein metric. The singularity at the tip of the cone can be smoothed, while preserving the Calabi--Yau structure, either by a small resolution or via a deformation, leading to the resolved or the deformed conifold respectively. These geometries can be used to construct supersymmetric solutions of type IIB supergravity whose dual field theories display some extremely interesting physical properties, like the breaking of conformality \cite{KlebanovNekrasov, KlebanovTseytlin, PandoZayasTseytlin}, confinement and chiral symmetry breaking \cite{KlebanovStrassler, MaldacenaNunez}. Recently, the fruitful interplay between the conifold and type IIB supergravity has been further investigated in \cite{MaldacenaMartelli}, where a solution exhibiting a geometric transition between the resolved and the deformed conifold purely within the supergravity framework was found.

The dual 4-dimensional field theories, and in particular their renormalization group flow, can be described holographically in a 5-dimensional setup, with the fifth (radial) coordinate playing the role of the renormalization group scale \cite{GPPZ, FGPW, GPPZ2}. The 10-dimensional solution can be reproduced in a 5-dimensional framework after having performed an appropriate dimensional reduction of type IIB supergravity on the compact transverse space ($T^{1,1}$ here). The conifold backgrounds mentioned above have just a radial profile; remarkably, in these cases the system of first-order differential equations for the supersymmetric radial flow generating the supergravity solution can be deduced from a single function of the fields, the superpotential \cite{KlebanovTseytlin, PandoZayasTseytlin, PapadopoulosTseytlin}. In particular, Papadopoulos and Tseytlin \cite{PapadopoulosTseytlin} obtained a truncation containing all the solutions in \cite{KlebanovTseytlin, PandoZayasTseytlin, KlebanovStrassler, MaldacenaNunez}, and providing the corresponding superpotentials. The truncation ansatz of \cite{PapadopoulosTseytlin} provides the metric plus a set of scalars in 5 dimensions. Under the further assumption of 4-dimensional Poincar\'e invariance, type IIB supergravity actually reduces to a 1-dimensional action describing the evolution of the fields with respect to the radial coordinate. 

In \cite{BergHaackMueck}, the Papadopoulos--Tseytlin ansatz was generalized to allow an arbitrary field dependence on all the spacetime coordinates transverse to the compact $T^{1,1}$ manifold, and the corresponding 5-dimensional action was derived. This extension allowed to extract informations about mass spectra and correlation functions of the dual field theory by studying supergravity field fluctuations in the 5-dimensional effective model.
As the truncation was shown to be consistent (modulo a Hamiltonian constraint to be imposed separately), any solution found in the lower-dimensional setup is guaranteed to lift to the full type IIB theory.

The Papadopoulos--Tseytlin ansatz has also been used to find a solution of type IIB supergravity interpolating between the Klebanov--Strassler \cite{KlebanovStrassler} and the Maldacena--Nu\~nez \cite{MaldacenaNunez} solutions \cite{BaryonicBranchKS}. Although these backgrounds, as well as those in \cite{KlebanovTseytlin, PandoZayasTseytlin} are supersymmetric, the effective 5-dimensional model of \cite{BergHaackMueck} does not exhibit manifest supersymmetry (even in the weaker sense that it does not coincide with the bosonic sector of a supersymmetric theory), and it was then classified as a fake supergravity model \cite{FakeSugra}. As a consequence, studying the supersymmetry properties of the solutions directly in 5 dimensions is not straightforward.

In this paper we overcome this limitation by embedding the models of \cite{PapadopoulosTseytlin, BergHaackMueck} in a genuine supergravity action. The latter is obtained via a consistent truncation of type IIB supergravity on the $T^{1,1}$ coset preserving $\mathcal N=4$ supersymmetry in 5 dimensions. 

In a previous work \cite{IIBonSE}, we derived a consistent truncation of type IIB supergravity on general squashed Sasaki--Einstein manifolds, leading to a gauged $\mathcal N=4$ supergravity in 5 dimensions with two vector multiplets (see also \cite{LiuEtAl, GauntlettVarelaSE, SkenderisTaylorTsimpis} for related work, and \cite{BuchelLiu, GauntlettVarela07, MaldacenaMartelliTachikawa, GauntlettKimVarelaWaldram, Gubser1} for previous consistent truncations on (squashed) Sasaki--Einstein manifolds). The truncation was based on expanding the 10-dimensional fields in a set of differential forms characterizing a Sasaki--Einstein structure. In the present paper we show that, for the specific case in which the compact manifold admitting a Sasaki--Einstein structure is the $T^{1,1}$ coset space, it is possible to enhance the consistent truncation of \cite{IIBonSE} by incorporating an additional vector multiplet of 5-dimensional $\mathcal N=4$ supergravity. This is obtained by adopting a truncation ansatz which retains all and only those modes of type IIB supergravity that are invariant under the SU(2)$\times$SU(2) acting transitively on the $T^{1,1}$ coset from the left. This symmetry protecting the retained modes guarantees the truncation to be consistent, while the preserved amount of supersymmetry is rather related to the structure group on $T^{1,1}$ being contained in SU(2).\footnote{In the slightly different context of $\mathcal N=2$ flux compactifications on SU(3) structure manifolds, an explicit proof of consistency for left-invariant reductions on coset spaces was given in \cite{ExploitingN=2}.}

In addition to the expansion forms considered in \cite{IIBonSE}, namely the minimal basis existing on any 5-dimensional manifold admitting a Sasaki--Einstein structure, the left-invariant forms on $T^{1,1}$ include a 2-form and a 3-form non-trivial in cohomology, reflecting the fact that topologically  $T^{1,1}\sim S^2\times S^3$. It follows in particular that, beside the RR 5-form flux considered in \cite{IIBonSE}, we can now include in the truncation both NSNS and RR 3-form fluxes. These lead to additional gaugings of the 5-dimensional supergravity model, that we analyse in detail applying the embedding tensor formalism. 

Together with the inclusion of the 3-form fluxes, the $\mathcal N=4$ vector multiplet that we are adding to the truncation of \cite{IIBonSE} is crucial for the gauge/gravity applications. Indeed, among the scalars contained in this multiplet there are both the resolution and the deformation modes of the conifold, as well as the modes of the type IIB 2-form potentials responsible for the running of the coupling constants in the dual field theories. Moreover, the new vector field is dual to the baryonic current operator. To make the dual picture of our truncation precise, we identify the full set of operators in the conifold gauge theory \cite{KlebanovWitten} which are dual to the supergravity modes we retain. This is done via a study of the mass spectrum about the supersymmetric AdS$_5$ background and a comparison with the results of~\cite{T11spectrum}.

As a first application of our consistent truncation, we search for new AdS$_5$ solutions of type IIB supergravity by extremizing the 5-dimensional scalar potential. We find an interesting family of non-supersymmetric backgrounds interpolating between a solution found long ago by Romans \cite{RomansNewIIBsol} and a new solution of the Freund--Rubin type involving an Einstein metric on $T^{1,1}$ not related to the usual Sasaki--Einstein one. However, a stability test against the left-invariant modes reveals that at least a subsector of our family of solutions is unstable.

Beside the non-supersymmetric model arising from the Papadopoulos-Tseytlin ansatz \cite{PapadopoulosTseytlin, BergHaackMueck}, and in addition to the cases already discussed in \cite{IIBonSE} (see also \cite{LiuEtAl, GauntlettVarelaSE, SkenderisTaylorTsimpis}), the present $\mathcal N=4$ reduction admits some further consistent truncations, which preserve just $\mathcal N=2$ supersymmetry and have a rather simple field content. As we will see, one of these truncations has the proper field content to capture the Klebanov--Strassler solution \cite{KlebanovStrassler}, and even to describe smeared D7-branes in this background \cite{Ramallo}. Moreover, a second $\mathcal N=2$ truncation provides a minimal supersymmetric completion to the one recently elaborated in \cite{HerzKlebPufuTesi} to study charged black 3-brane solutions possibly relevant for the holographic description of condensed matter phenomena. We hope that our supersymmetric truncations could also be applied to properly embed into string theory models describing condensed matter systems.

The paper is organized as follows. In section \ref{T11CosetGeom} we introduce the $T^{1,1}$ geometry, focusing on its coset structure and computing the left-invariant tensors. In section \ref{DimRed} we elaborate our SU(2)$\times$SU(2) invariant truncation ansatz for type IIB supergravity, and present the 5-dimensional action. The compatibility of this model with the structure of gauged $\mathcal N=4$ supergravity is detailed in section \ref{gaugedsugra}. We continue in section \ref{NewFamilyAdS} by presenting our new interpolating family of AdS$_5$ solutions, while in section \ref{DualOperators} we identify the dual operators. The consistent subtruncations preserving $\mathcal N=2$ supersymmetry are discussed in section \ref{N=2subtrunc}. We conclude in section \ref{conclusions} by providing some possible lines of future research. Three appendices complete the paper:
appendix \ref{Reduct10dRicci} contains the reduction of the 10-dimensional Ricci tensor,
appendix \ref{RecoveringPT} provides the dictionary between our truncation and the Papadopoulos--Tseytlin one, and appendix \ref{GaugeTransf} discusses the 5-dimensional gauge transformations arising from a reduction of the 10-dimensional symmetries.

\vskip 3mm

\noindent {\bf Note added}: On the same day this paper appeared on the arXiv, the work \cite{Bena:2010pr} also appeared, which has considerable overlap with our sections \ref{DimRed} and \ref{gaugedsugra}.

\section{The coset geometry}\label{T11CosetGeom}

In this section we introduce the (left) coset 
\begin{equation} T^{1,1}= \frac{{\rm SU}(2)\times {\rm SU}(2)}{{\rm U}(1)}\,,
\end{equation}
where U(1) is embedded diagonally in SU(2)$\times$SU(2). Topologically this is $S^2\times S^3$, and can also be seen as a U(1) fibration over $S^2\times S^2\,$ \cite{CandelasOssaConifolds}. In the following we provide the most general left-invariant metric and the set of left-invariant differential forms. These will be the building blocks of our truncation ansatz, to be implemented in the next sections. For details on the general theory of coset manifolds we refer to e.g. \cite{MuellerStuckl, CastDAuriaFreBook}.

We choose SU(2)$\times$SU(2) generators
\begin{eqnarray}
T_{1,2} \!\!\!&=&\!\!\! \frac{1}{2i}\sigma_{1,2}\,,\qquad T_{3,4} = \frac{1}{2i}\tilde\sigma_{1,2}\,,\qquad T_5 = \frac{1}{4i}(\sigma_3 - \tilde\sigma_3)\,,\qquad T_6 = \frac{1}{4i}(\sigma_3 + \tilde\sigma_3)\,,
\end{eqnarray}
where $\{\sigma_1,\sigma_2,\sigma_3\}$ and $\{\tilde\sigma_1,\tilde\sigma_2,\tilde\sigma_3\}$ are Pauli matrices for the first and second SU(2) factor respectively. The $T^{1,1}$ coset is defined by modding out with respect to the U(1) generated by $T_6$. The structure constants associated with the generators above are
\begin{eqnarray}
\nonumber c^5{}_{12}\!\!&=&\!\! c^6{}_{12} = -c^5{}_{34} = c^6{}_{34} = 1\,,\\ [2mm]
c^1{}_{25} \!\!&=&\!\! - c^2{}_{15} = c^4{}_{35} = - c^3{}_{45} = c^1{}_{26} =  -c^2{}_{16} = c^3{}_{46}  = -c^4{}_{36} = \frac{1}{2}\:.
\end{eqnarray}
Then on the coset there exist local coframe 1-forms $\{e^a\}$, $a=1,\ldots, 5\,$, satisfying
\begin{equation}\label{eq:ExtDerCoframe}
de^a\, =\, - \frac{1}{2}c^{a}{}_{bc}e^b\wedge e^c - c^{a}{}_{b6}e^b\wedge e^6\,,
\end{equation}
where the 1-form $e^6$ is a connection term associated with the U(1) by which we mod out.
An explicit expression for the $e^a$ in terms of angular coordinates is given in appendix \ref{RecoveringPT}.

The left-invariant tensors on $T^{1,1}$ are those being invariant under the left-action of the group SU(2)$\times$SU(2) on the coset itself. It follows that such tensors are globally defined. For the metric
\begin{equation}
ds^2(T^{1,1})\,=\, g_{ab}\,e^a\otimes e^b\,,\qquad\quad a,b=1,\ldots, 5\,,
\end{equation} 
the left-invariance condition requires that its coframe components $g_{ab}$ be independent of the coset coordinates, and satisfy
\begin{equation}
c^c{}_{6(a} \,g_{b)c} = 0\,.
\end{equation}
Similarly, a differential $p$-form $\varphi= \frac{1}{p!}\varphi_{a_1\ldots a_p}e^{a_1}\wedge\ldots \wedge e^{a_p}$ is left-invariant if its components are constant over the coset and
\begin{equation}
c^b{}_{6[a_1} \varphi_{a_2 \ldots a_p]b } = 0\,.
\end{equation}
A relevant property of left-invariant forms is that their exterior derivative is still left-invariant, with the connection term in (\ref{eq:ExtDerCoframe}) always dropping out.

\subsection{Metric and curvature}\label{T11MetricCurvature}

The most general left-invariant metric on $T^{1,1}$ depends on five arbitrary parameters:
\begin{equation}\label{GenericLeftInvMetric}
g_{ab} \;=\; \left(\begin{array}{cc|cc|c} 
A \!&\!  \!&\! D \!&\! E &\! \\  [.3mm]
 \!&\! A \!&\! -E \!&\! D &\! \\  [.3mm] \hline
\rule{0pt}{2.3ex} D \!&\! -E \!&\! B \!&\!  &\! \\ [.3mm]
E \!&\! D \!&\!  \!&\! B &\! \\  [.3mm]\hline \rule{0pt}{2.3ex} 
 \!&\!  \!&\!  \!&\!  &\! C
\end{array}\right).
\end{equation}
We will adopt the following convenient reparameterization\footnote{Throughout the paper, we use the space-saving notation $\ch \equiv \cosh$ and $\sh \equiv\sinh $.}
\begin{eqnarray}
\nonumber A  \!&\!=\!&\! \frac{1}{6}\,e^{2u+2w}\,\ch\, t\;,\qquad \qquad B\,=\, \frac{1}{6}\, e^{2u-2w}\,\ch\, t \;,\qquad\qquad C \,=\, \frac{1}{9}\,e^{2v}\\ [2mm]
D \!&\!=\!&\! \frac{1}{6}\,e^{2u}\,\sh\, t\,\cos\theta \;,\,\quad\qquad E\,=\, \frac{1}{6}\,e^{2u}\,\sh\, t\,\sin\theta  \;, \qquad\quad  \;\; t \geq 0\,,\label{eq:ReparamMetric}
\end{eqnarray}
which also ensures metric positivity. The choice of the numerical multiplicative factors sets the Sasaki--Einstein point at the origin of the metric parameter space:
\begin{equation}\label{eq:SEpoint}
\textrm{Sasaki--Einstein metric}\qquad\Leftrightarrow\qquad u\,=\, v\,=\,w \,=\,t = 0\,.
\end{equation}
In the $e^a$ coframe, the Ricci tensor $R_{ab}$ has the same structure as the metric (\ref{GenericLeftInvMetric}), and reads
\begin{eqnarray}
\nonumber R_{11} \!&=&\! 1+ \ch\, t\, \Big[-\frac{1}{3} e^{-2 (u-v+w)}+\frac{3}{4} e^{2
(u-v+w)} \shsh t\,\Big]\,,\qquad \quad R_{33}\;=\; R_{11}|_{w\,\to\, -w}\;,\\[2mm]
\nonumber R_{55}\!&=&\! \frac{4}{9}\, e^{-4( u- v)} \Big[\,\chch t\; {\rm ch}(4 w)- \shsh t \,\Big] - \shsh t \,,\\[2mm]
R_{13} \!&=&\! \sh\, t \,\Big[-\frac{1}{3} e^{-2 (u-v)}+\frac{3}{4} e^{2 (u-v)}\chch t\,\Big] \cos\theta\,,\qquad \qquad\;\;    R_{14} \;=\; R_{13}\tan\theta \;.\qquad\,
\end{eqnarray}
Then the Ricci scalar is
\begin{equation}\label{R_T11}
R_{T^{1,1}} \,=\, 4\,e^{-4 u+2v} \left[ \shsh t -\chch t\; {\rm ch}(4 w)\right]  + 24\, e^{-2 u}\, \ch\, t \;{\rm ch}(2 w) -9 \,e^{-2 v} \shsh t \,.
\end{equation}

The parameters $u$ and $v$ describe the ``breathing'' and ``squashing'' modes of the $T^{1,1}$ metric: indeed, $4u+v$ is the overall volume, while $u-v$ modifies the relative size of the \hbox{4-dimensional} base and the U(1) fibre when $T^{1,1}$ is seen as a U(1) fibration over $S^2\times S^2$. Moreover, the parameter $w$ controls the relative size of the two $S^2$s, and breaks their exchange symmetry.
Finally, it is interesting to notice that the left-invariant metric above is suitable for describing both the deformed and the resolved conifolds. In particular, the parameter $t$ enters in the description of the former, while $w$ plays a role in the latter (explicit expressions for the respective metrics can be found e.g. in \cite{MinasianTsimpis, PandoZayasTseytlin, PapadopoulosTseytlin}).

\subsection{Left-invariant forms}\label{LeftInvForms}

The left-invariant forms on the $T^{1,1}$ coset are spanned by the 1-form $e^5$ and the four 2-forms $ e^{12},\;  e^{34} ,\; e^{13} + e^{24}, \; e^{14}- e^{23}\,,$ together with all their possible wedgings. We combine them into the following equivalent basis:
\begin{eqnarray}
\nonumber \eta &=&  -\frac{1}{3}e^5 
\,,\qquad\qquad \qquad\qquad \quad\Omega \;=\; \frac{1}{6}(e^1+i e^2)\wedge (e^3-ie^4)\,, \\ [3mm]
J &=& \frac{1}{6}(e^{12} - e^{34}) 
\;,\qquad \qquad \qquad\Phi \;=\; \frac{1}{6}(e^{12} + e^{34})\,.\label{leftinvforms}
\end{eqnarray}
These satisfy the algebraic conditions 
\begin{eqnarray}
\nonumber\eta\lrcorner J \;=\; \eta\lrcorner \Omega \;=\; \eta\lrcorner \Phi &=& 0\\ [2mm]
\nonumber\Omega\wedge \Omega \;=\; \Omega\wedge J \;=\; \Omega\wedge \Phi \;=\; J\wedge \Phi &=& 0\\ [2mm]
\Omega\wedge \overline\Omega \;=\; 2 J\wedge J \;=\; -2\Phi\wedge \Phi &\neq& 0\,,
\end{eqnarray}
defining a reduction of the structure group on the coset manifold to U(1)$\,\subset\,$SU(2)$\,\subset\,$SO(5). Furthermore, the following differential relations are satisfied
\begin{equation}
d\eta = 2J\;,\quad\qquad 	d \Omega \;=\; 3 i\, \eta \wedge \Omega\,, \quad\qquad  d\Phi = 0  \label{SEstructure}\,.
\end{equation}
This implies that the forms $\eta,\,J, \,\Omega$ characterize a Sasaki--Einstein structure. These were employed in \cite{IIBonSE} as a basis to expand the 10-dimensional fields and derive a consistent truncation of type IIB supergravity on general squashed Sasaki--Einstein manifolds. Here, the non-trivial cohomology of $T^{1,1}\sim S^2\times S^3$ leads us to consider the additional  form $\Phi$. Indeed, while $J$ is exact and $\Omega$ is non-closed, $\Phi$ and $\Phi\wedge \eta$ respectively span the second and third cohomology of $T^{1,1}$.

For the dimensional reduction to go through, it is also crucial to remark that left-invariance ensures the closure of the set of basis forms under the 5-dimensional Hodge star operation. In detail we find
\begin{eqnarray}\label{eq:T11HodgeStar}
\nonumber *1 \!\!&=&\!\! \frac{1}{2} e^{4u+v}  J\wedge J \wedge \eta\,, \\ [2mm]
\nonumber *\eta \!\!&=&\!\! \frac{1}{2} e^{4u-v} J\wedge J\,, \\ [2mm]
\nonumber *J \!\!&=&\!\! e^v\left[\,\big(\chch t\:\ch(4w)-\shsh t\big)J + \chch t \:\sh(4w)\Phi -\sh(2t)\,\sh(2w)\, {\rm Im}( e^{i\theta}\Omega)\,\right]\wedge \eta\,,\\ [2mm]
\nonumber *\Phi \!\!&=&\!\! e^v\left[\,-\chch t \,\sh(4w)J - \big(\chch t \:\ch(4w) + \shsh t\big)\Phi + \sh(2t)\,\ch(2w)\, {\rm Im}(e^{i\theta}\Omega)\,\right]\wedge \eta\,,\\ [2mm]
*\Omega \!\!&=&\!\!  e^v\left[\,\chch t \,\Omega - \shsh t \, e^{-2i\theta}\,\overline\Omega - i e^{-i\theta} \sh(2t) \,\big(\,\sh (2w) J + \ch(2w)\Phi\,\big)\, \right]\wedge \eta \,.
\end{eqnarray}

\section{The dimensional reduction}\label{DimRed}

In this section we spell out our SU(2)$\times$SU(2) invariant ansatz for truncating type IIB supergravity. Next we implement the dimensional reduction, and present the resulting \hbox{5-dimensional model}. While the procedure follows quite closely the one implemented in \cite{IIBonSE}, here we will try to emphasize the points of difference.

The type IIB supergravity action is\footnote{In this paper, for any $p$-form $\varphi$ we use the shorthand notation $\varphi^2\equiv\varphi\,\lrcorner\,\varphi$, with the index contraction~$\lrcorner$ including the $\frac{1}{p!}$ factor.} 
\begin{eqnarray}
\nonumber S_{\rm IIB} \!\!&=&\!\! \frac{1}{2\kappa_{10}^2}\int \Big[ R - \frac{1}{2}(d\phi)^2 - \frac{1}{2}e^{-\phi} H^2 - \frac{1}{2}e^{2\phi} (F_1)^2 - \frac{1}{2} e^{\phi} (F_3)^2- \frac{1}{4} (F_5)^2 \Big] *1\\ [2mm]
\label{eq:actionIIB} &-&\!\!\frac{1}{8\kappa_{10}^2}\int  (B\wedge dC_2-C_2\wedge dB)\wedge dC_4\,,
\end{eqnarray}
where the form field-strengths need to satisfy the Bianchi identities
\begin{equation}\label{eq:IIBbianchis}
dH=0\,,\qquad\; dF_1=0 \,,\qquad \; dF_3 = H\wedge F_1\,, \qquad \; dF_5 = H\wedge F_3\,,
\end{equation}
to be solved in terms of the NSNS potential $B$ and RR potentials $C_0,\,C_2,\,C_4$. In this regard, there are some subtleties due to the presence of background fluxes that will be discussed below. Furthermore, the RR 5-form has to satisfy the self-duality constraint $F_5=*F_5\,$.

\subsection{Reduction of the curvature}\label{Red10dCurvature}

We take the 10-dimensional spacetime to be a direct product $M\times T^{1,1}$, where $M$ is the 5-dimensional spacetime. Our ansatz for the 10-dimensional metric in the Einstein frame is 
\begin{eqnarray}
\nonumber d s^2 &=& e^{-\frac{2}{3}(4u+v)} g_{\mu\nu}dx^\mu dx^\nu + g_{ab}(e^a -\delta^a_5 \,3A  ) (e^b -  \delta^b_5\, 3A)\\ [3mm]
&=& e^{-\frac{2}{3}(4u+v)} g_{\mu\nu}dx^\mu dx^\nu \,+\, \frac{1}{6}e^{2u}\ch\,t \left[  e^{2w} (e^1e^1+e^2e^2) + e^{-2w}(e^3e^3+e^4e^4)\right] \nonumber \\ [2mm]
&+&  e^{2v}(\eta + A)^2 \,+\, \frac{1}{3}e^{2u}\sh\,t \left [ \cos\theta (e^1e^3 +e^2e^4) + \sin\theta  (e^1e^4-e^2e^3) \right],\label{10dmetric}
\end{eqnarray}
where $x^\mu$ are coordinates on $M$, whose metric is $g_{\mu\nu}(x)$. The dependence on the  $T^{1,1}$ coordinates is relegated to the coset 1-forms $\{e^a\}_{a=1,\ldots, 5}$ introduced above, and in this frame the metric $g_{ab}$ was given in (\ref{GenericLeftInvMetric}), (\ref{eq:ReparamMetric}); now its five parameters $u,v,w,t,\theta$ are promoted to scalar fields on $M$. Finally, $A(x)$ is a 1-form on $M$.

The transformation 
\begin{equation}\label{eq:Shiftpsi}
\psi\to\psi -3\,\omega(x)
\end{equation} 
of the $T^{1,1}$ coordinate $\psi$ entering in $\eta \equiv -\frac{1}{3}e^5 = -\frac{1}{3}d\psi + \ldots\;$ (cf. eq. (\ref{vielbeins})) is interpreted as an abelian gauge transformation for the 1-form $A$, which is shifted as $A \to A+d\omega$. The field theory dual of this symmetry is the U(1) R-symmetry. Furthermore, by looking at the explicit coordinate expression of the $T^{1,1}$ metric (given by plugging (\ref{vielbeins}) into (\ref{10dmetric})), one can see that the coordinate $\psi$ and the metric parameter $\theta$ always appear in the combination $\psi - \theta$, hence the transformation (\ref{eq:Shiftpsi}) also shifts $\theta$, which at the 5-dimensional level is then interpreted as the phase of a charged scalar.

Using the reduction formulae for the curvature provided in appendix~\ref{Reduct10dRicci}, we obtain that the 10-dimensional Einstein--Hilbert term reduces to the 5-dimensional action
\begin{eqnarray}\label{S_PureMetric0}
\frac{1}{2\kappa_{10}^2}\int (R*\!1)_{10} \!\!&=&\!\! \frac{1}{2\kappa_5^2}\int \Big[ R \,-\, \frac{1}{2}e^{\frac{8}{3}u+\frac{8}{3}v} (dA)^2\, +\, e^{-\frac{8}{3}u-\frac{2}{3}v}R_{T^{1,1}} \\ [2mm]
\nonumber&&\!\! -\frac{28}{3}du^2 -\frac{4}{3}dv^2 - \frac{8}{3}du\lrcorner dv - dt^2 - 4\,\chch t \,dw^2 - \shsh t\,(d\theta - 3A)^2 \Big]\!*\!1.
\end{eqnarray}
where the internal Ricci curvature $R_{T^{1,1}}$, providing a scalar potential term, was given in (\ref{R_T11}), while the 5-dimensional gravitational coupling $\kappa_5^2$ is defined in (\ref{eq:5dcoupling}).

The fact that the internal metric modes provide a charged scalar is best seen by considering the redefined fields
\begin{equation}
e^{2\tilde w} = e^{2w}\ch\, t\;,\qquad\qquad\tilde t = e^{-2w}\tanh t\, ,
\end{equation}
so that the scalar kinetic terms become
\begin{equation}
- dt^2 - 4\,\chch t\,dw^2  - \shsh t\,(d\theta - 3A)^2 \;\,=\,\; -4\, d\tilde w^2 -e^{4\tilde w}|D z|^2\,,
\end{equation}
where the complex scalar $z = \tilde t\, e^{i\theta}$ is charged under the gauge potential $A$, with gauge covariant derivative
$\,Dz \,=\, (d-3iA)z\,$. It follows that whenever $t$ takes a non-trivial vacuum expectation value, the U(1) gauge symmetry is spontaneously broken and $A$ acquires a mass via the Higgs mechanism. We will see an example of this in section~\ref{NewFamilyAdS}.

Finally, we remark that in the limit $w=0,\,t=0\,$, the metric ansatz adopted in \cite{IIBonSE} is recovered, as well as the corresponding curvature.\footnote{In \cite{IIBonSE}, the scalars $u$ and $v$ appearing here were called $U$ and $V$ respectively.}

\subsection{Expansion of the form fields}\label{FormFields}

We now pass to define the truncation ansatz for the remaining type IIB supergravity fields. The prescription is to write down the most general expansion compatible with SU(2)$\times$SU(2) invariance; this is implemented by using the basis forms introduced in section \ref{LeftInvForms}.

The dilaton and the RR axion are assumed independent of the internal coordinates: $\phi = \phi(x)\,,$ $C_0 = C_0(x)\,$, and $F_1=dC_0$. For the NSNS 3-form we take
\begin{equation}\label{eq:10dExpH}
H = H^{\rm fl} + dB\,,
\end{equation}
where the flux piece is given in terms of the cohomologically non-trivial 3-form on $T^{1,1}$ as
\begin{equation}
H^{\rm fl}= p \,\Phi\wedge \eta\,,    \qquad\qquad p= {\rm const}\,,
\end{equation}
and the potential is expanded in the left-invariant forms (\ref{leftinvforms}) as
\begin{equation}
	\label{eq:Bcov}	B = b_2 + b_1 \wedge(\eta + A) + b^{J} J + {\rm Re}(b^\Omega\,\Omega) + b^\Phi \Phi\,,
\end{equation}
where $b_p \equiv b_p(x)$ are $p$--forms on $M$ (we omit the 0 subscript for the scalar fields). The field labeled with $\Omega$ is complex, while the others are real. We are not introducing the potential $B^{\rm fl}$ giving rise to $H^{\rm fl}$ because, in contrast to the latter, it is not globally defined and therefore not left-invariant: in terms of the $\psi$ coordinate on $T^{1,1}$ introduced in appendix~\ref{RecoveringPT}, locally this can be written as $B^{\rm fl}= - \frac{p}{3}\,\psi\,\Phi\,$.
The expansion of $H$ is
\begin{eqnarray}\label{eq:ExpH}
H&=&h_3+h_2\wedge\left(\eta+A\right)+h_1^J\wedge J+ {\rm Re}\big[ h_1^\Omega\wedge\Omega+ h_0^{\Omega}\, \Omega\wedge(\eta + A)\big]\nonumber\\[2mm]
&&+\, h_1^\Phi\wedge\Phi+p\,\Phi\wedge\left(\eta+A\right)\,,
\end{eqnarray}
where, recalling (\ref{SEstructure}), the $h_p$ read
\begin{equation}
\begin{array}{rclcrcl}
h_3 &=& db_2 - b_1\wedge dA\,,\;\;&&\;\;h_1^\Omega &=& db^\Omega-3iA\,b^\Omega \,\equiv\, Db^\Omega, \\[3mm]
h_2 &=& db_1\,, \;\;&&\;\; h_0^\Omega &=& 3ib^\Omega, \\[3mm]
h_1^J &=& db^J -2b_1 \,\equiv\, Db^J \,,\quad&&\quad h_1^\Phi &=& db^\Phi-p\,A \,\equiv\,  Db^\Phi \,.
\end{array}
\end{equation}
As far as the terms involving the forms $\{\eta,J,\Omega\}$ are concerned, the expansions above are the same as those performed in \cite{IIBonSE}, where we described a universal truncation of type IIB supergravity on squashed Sasaki--Einstein manifolds. The cohomologically non-trivial forms $\Phi$ and $\Phi\wedge \eta$ we are considering in this work allow to introduce an additional scalar, $b^\Phi$, together with a flux term, $p$. Note that from the 5-dimensional perspective, the effect of having non-closed expansion forms on the internal manifold (namely, $\eta$ and $\Omega$) is similar to the effect of the NSNS and RR fluxes: they induce non-trivial derivatives for the scalars. This justifies why the coefficients in the expansion of the exterior derivative of a basis form are sometimes called geometric fluxes. Moreover, this also shows that the supergravity model we are going to obtain in  5-dimensions has to be gauged: due the fluxes, would-be neutral scalar fields become charged under some of the vectors present in the model. Some scalars are properly charged under a U(1) (like the complex scalar $b^\Omega$), while some others are rather axions having St\"uckelberg couplings (like $b^J,\,b^\Phi$). More details about the gauge structure of the 5-dimensional theory are given in section \ref{gaugedsugra} and in appendix \ref{GaugeTransf}.

The expansion of the RR 3-form
\begin{equation}\label{eq:10dExpF3}
F_3 \,=\, F_3^{\rm fl} + dC_2 - C_0\, H
\end{equation}
proceeds in a similar way, with 
\begin{equation}
F_3^{\rm fl}=q\,\Phi\wedge\eta\,,    \qquad\qquad q = {\rm const}\,,
\end{equation} 
and with new 5-dimensional forms $c_p$, $g_p$ replacing $b_p$, $h_p$ in the expansion of $C_2$, $F_3$ respectively. In this case the identifications are
\begin{equation}
\begin{array}{rclcrcl}
g_3 &=& dc_2 - c_1\wedge dA-C_0(db_2 - b_1\wedge dA)\,,\quad&&\quad g_1^\Omega &=&Dc^\Omega-C_0 Db^\Omega, \\[3mm]
g_2 &=&dc_1-C_0 db_1\,, \;\;&&\;\; g_0^\Omega &=& 3i\left(c^\Omega-C_0b^\Omega\right), \\[3mm]
g_1^J &=& Dc^J-C_0Db^J \,,\;\;&&\;\; g_1^\Phi &=& Dc^\Phi-C_0Db^\Phi,
\end{array}
\end{equation}
where $D c^J$ and $Dc^\Omega$ are defined analogously to $D b^J$ and $Db^\Omega$, with the replacement $b\to c$, while $Dc^\Phi \equiv dc^\Phi-q\,A$.

Passing to the RR 5-form, we solve its Bianchi identity (\ref{eq:IIBbianchis}) by
\begin{equation}\label{eq:10dExpF5}
F_5 \;=\; F_5^{\rm fl} + dC_4 + \frac{1}{2}\big[B\wedge (dC_2 + 2F_3^{\rm fl}) - C_2 \wedge (dB + 2H^{\rm fl} )\big],
\end{equation}
which has the advantage of involving globally defined forms only, that can be directly expanded in a left-invariant basis. Assuming the expansion 
\begin{eqnarray}
F_5 \!\!&=&\!\! f_5+f_4\wedge\left(\eta+A\right)+f_3^J\wedge J+ f_2^J\wedge J\wedge\left(\eta+A\right) + {\rm Re}\left[f_3^\Omega\wedge\Omega + f_2^\Omega\wedge\Omega\wedge\left(\eta+A\right)\right] \nonumber\\ [3mm]
\!\!&&\!\!+\; f_3^\Phi\wedge\Phi+f_2^\Phi\wedge\Phi\wedge\left(\eta+A\right)+ f_1\wedge J\wedge J+f_0\,J\wedge J\wedge\left(\eta+A\right) ,\label{eq:ExpF5}
\end{eqnarray}
from (\ref{eq:10dExpF5}) we can identify the expression for all the 5-dimensional $p$-forms $f_p\,$. However, we also need to eliminate the redundancy in the degrees of freedom of $F_5$ by imposing the self-duality condition. This translates into a series of 5-dimensional relations, listed in eq. (\ref{5dselfduality}) of the appendix, that allow to express $f_5,\,f_4,\,f_3^J,\,f_3^\Omega,\,f_3^\Phi$ in terms of $f_0,\,f_1,\,f_2^J,\,f_2^\Omega,\,f_2^\Phi$. The latter fields thus encode the independent degrees of freedom carried by $F_5$, and read
\begin{eqnarray}
f_0&=&3\,{\rm Im}(b^\Omega\overline{c^\Omega})+p\,c^\Phi-q\,b^\Phi+k\nonumber\\[3mm]
f_1&=&Da+\frac{1}{2}\left(q\,b^\Phi-p\,c^\Phi\right)A+\frac{1}{2}\left[b^J\,Dc^J-b^\Phi\,Dc^\Phi+{\rm Re}(b^\Omega\,\overline{Dc^\Omega})-\,\,b\,\leftrightarrow\,c\right]\nonumber\\[3mm]
\nonumber f_2^J &=& da_1^J  +\frac{1}{2}\left[\,b^Jdc_1-b_1\wedge D c^J - \,\,b\leftrightarrow c\,\right], \\[3mm]
\nonumber f_2^\Omega &=& Da_1^\Omega+3ia_2^\Omega+\frac{1}{2}\left[b^\Omega dc_1-b_1\wedge Dc^\Omega+3ic^\Omega b_2- \,\,b\leftrightarrow c\right],	\\[3mm]
f_2^\Phi&=&da_1^\Phi+\frac{1}{2}\left(q\,b_1-p\,c_1\right)\wedge A+q\,b_2-p\,c_2+\frac{1}{2}\left[b^\Phi\,dc_1-b_1\wedge Dc^\Phi-\,\,b\,\leftrightarrow\,c\right]\qquad\label{eq:Expf}
\end{eqnarray}
where the $a_p$ fields come from the expansion of $C_4$. Moreover, $Da=da-2a_1^J - kA$, while $a_1^\Omega$ should be regarded as a pure gauge field for the complex 2-form $a_2^\Omega$ (see \cite{IIBonSE} for more details).

Having completed the truncation ansatz for the type IIB supergravity fields, we can proceed with the dimensional reduction of the corresponding terms in the action (\ref{eq:actionIIB}). This simply requires to plug their expansion in the action, use the relations among the basis forms, and take the integral over the internal manifold. In doing this, there are however two issues that need to be considered.

The first is that, due to the presence of background fluxes, the full $B,\, C_2$ and $C_4$ potentials appearing in the standard topological term of the type IIB action (\ref{eq:actionIIB}) are only locally defined. On the other hand, in order to expand in our left-invariant basis, we require a formulation involving just globally defined forms. This already motivated the quite non-standard definition (\ref{eq:10dExpF5}) of the $F_5$ field-strength. Such requirement can be fulfilled at the expense of modifying the topological term.\footnote{Similar conclusions were reached in the context of type IIA with fluxes in \cite{KachruKashaniPoor, IIAModuliStabilization, ReducingSU3SU3}.} By imposing that the variation with respect to the globally defined potentials still give the correct type IIB equations of motion, we find that the proper form of the topological term, replacing the second line in (\ref{eq:actionIIB}), is
\begin{eqnarray}
\nonumber S_{\rm IIB,\,top} \!\!&=&\!\! -\,\frac{1}{8\kappa_{10}^2}\int \Big[ \Big(B\wedge (dC_2+2F_3^{\rm fl}) - C_2\wedge (dB+2H^{\rm fl})\Big)\wedge (dC_4+F_5^{\rm fl})\;\\ [2mm]
\label{IIBTopTermModified} &&  \qquad\qquad +\,\frac{1}{2}\big( B\wedge B \wedge dC_2\wedge F_3^{\rm fl} + C_2\wedge C_2 \wedge dB \wedge H^{\rm fl}  \big)  \Big],
\end{eqnarray}
where the potentials $B,\,C_2,\,C_4$ are now the ones appearing in (\ref{eq:10dExpH}), (\ref{eq:10dExpF3}), (\ref{eq:10dExpF5}). 

The second delicate point concerns the reduction of the terms involving the RR 5-form $F_5$, because of the self-duality condition which makes the $F_5$ kinetic term vanish on-shell. In order to implement this constraint, we follow the procedure illustrated in \cite{IIBonSE}.  The rest of the dimensional reduction goes through straightforwardly, and its outcome is presented below.

\renewcommand{\arraystretch}{1.2}
\begin{table}
\begin{center}
$\begin{array}{c|c|c|c|c}
\textrm{IIB fields}    & 	\textrm{scalars}   &  \textrm{1-forms}   & \textrm{2-forms} & \textrm{5d metric} \\
\hline
\textrm{10d metric} &  u,\,v,\,w,\,t,\,\theta & A  &  &  g_{\mu\nu}  \\ 
\phi & \phi  &  &  & \\ 
B & b^J,\,b^\Phi,\, b^\Omega  & b_1  & b_2  & \\ 
C_0 & C_0 &   &   & \\ 
C_2 & c^J,\,c^\Phi,\, c^\Omega  & c_1  & c_2  & \\
C_4 & a  &  a_1^J,\,a_1^\Phi,\,a_1^\Omega & a_2^\Omega  &
\end{array}$
\caption{Summary of the 5-dimensional bosonic fields and their type IIB supergravity origin. The fields labeled with $\Omega$ are complex, while all the others are real. The gauge field $A$ also appears in the expansion of the type IIB forms, see e.g. eq. (\ref{eq:Bcov}). The fluxes of the type IIB field strengths $F_5,\,F_3$ and $H$ are described respectively by the parameters $k$, $q$ and $p$.} \label{Summary5dfields}
\end{center}
\end{table}

\subsection{The five-dimensional model}\label{eq:5dmodel}

In the following we present the 5-dimensional model arising from the dimensional reduction on the $T^{1,1}$ coset. The consistency of the truncation is guaranteed by the SU(2)$\times$SU(2) invariance of our ansatz. The latter provides the field content summarized in table~\ref{Summary5dfields}. However, also in order to check the correctness of the 5-dimensional action below, we verified that the corresponding 5-dimensional equations of motion match the ones obtained by reducing the 10-dimensional type IIB equations. The 5-dimensional action takes the form
\begin{equation}
S\;=\;\frac{1}{2\kappa_5^2}\int R*1 \,\,+\,S_{\rm kin,scal} \,+\, S_{\rm kin,vect} \,+\, S_{\rm kin,forms} \,+\, S_{\rm top} \,+\,  S_{\rm pot}\,,
\end{equation}
where we have divided it into the gravitational part, the kinetic terms for forms of different degrees (respectively, scalars, 1-forms and 2-forms), the topological terms and finally the scalar potential.
The scalar kinetic terms are
\begin{eqnarray}\label{scalkinterms}
S_{\rm kin,scal}\!\!&=&\!\!-\frac{1}{2\kappa_5^2}\,\int\,\bigg\{\frac{28}{3}du^2 +\frac{4}{3}dv^2 + \frac{8}{3}du \lrcorner dv + dt^2 + 4\,\chch t\,dw^2 + \shsh t \,(d\theta - 3A)^2  \nonumber\\[2mm]
&&\qquad   +\, e^{-4u-\phi}\Big[\left(\chch t\,\ch(4w)-\shsh t\right)(h_1^J)^2+\left(\chch t\,\ch(4w)+\shsh t\right)(h_1^\Phi)^2\nonumber\\[3mm]
&& \qquad \qquad \qquad +\,\chch t \,|h_1^\Omega|^2- \shsh t\, {\rm Re}\left(e^{-2i\theta} (h_1^\Omega)^2\right)- 2\,\chch t \,\sh(4w)\,h_1^J\,\lrcorner\, h_1^\Phi\nonumber\\[3mm]
&& \qquad \qquad \qquad   -\,2 \,\sh(2t) \big(\sh(2w)h_1^J - \ch(2w)h_1^\Phi \big)\,\lrcorner\, {\rm Re}\big(i \,e^{-i\theta} h_1^\Omega \big) \Big]\nonumber\\[2mm]
&& \qquad   + e^{-4u+\phi}\Big[\,h\,\rightarrow\,g\,\Big]  \, +\, \frac{1}{2}d\phi^2 \,+\, \frac{1}{2}e^{2\phi}dC_0^2 \,+\, 2\,e^{-8u}\,f_1^2   \bigg\}*1\,,
\end{eqnarray}
where $[h\to g]$ stands for the repetition of the terms in the previous square bracket with $h$ replaced by $g$. The kinetic terms of the 1-forms are
\begin{eqnarray}
S_{\rm kin,vect} \!\!\!&=&\!\!\! -\frac{1}{2\kappa_5^2}\,\int\,\bigg\{\, \frac{1}{2}e^{\frac{8}{3}u+\frac{8}{3}v}\,(dA)^2 +\frac{1}{2}e^{\frac{8}{3}u-\frac{4}{3}v-\phi}\,h_2^2 +\frac{1}{2}e^{\frac{8}{3}u-\frac{4}{3}v+\phi}\,g_2^2\nonumber\\[2mm]
&&\quad  +\,e^{-\frac{4}{3}u-\frac{4}{3}v}\Big[\left(\chch t\,\ch(4w)-\shsh t\right)(f_2^J)^2+\left(\chch t\,\ch(4w)+\shsh t\right)(f_2^\Phi)^2\nonumber\\[3mm]
&&\qquad \qquad \quad -\, \shsh t\, {\rm Re}\left(e^{-2i\theta}\, (f_2^\Omega)^2\right)+\chch t \,|f_2^\Omega|^2- 2\,\chch t \,\sh(4w)\,f_2^J\lrcorner f_2^\Phi\nonumber\\[3mm]
&&\qquad \qquad \quad -\,2 \,\sh(2t) \big(\sh(2w)f_2^J - \ch(2w)f_2^\Phi \big)\,\lrcorner\,{\rm Re}\big(i \,e^{-i\theta} f_2^\Omega \big) \Big]\  \bigg\}*1\,.\qquad
\end{eqnarray}
The kinetic terms for the 2-forms read
\begin{equation}
S_{\rm kin,forms}\,=\,-\frac{1}{4\kappa_5^2}\,\int\, e^{\frac{16}{3}u+\frac{4}{3}v} \big(\, e^{-\phi}h_3^2 + e^{\phi} g_3^2\, \big)*1\, ,
\end{equation}
In addition we get some rather involved 5-dimensional topological couplings, which read
\begin{eqnarray}
\nonumber S_{\rm top} \!\!&=&\!\! \displaystyle\frac{1}{2\kappa_{5}^2}\int \left\{ \frac{i}{3}(\,\overline{Da_1^\Omega + 3ia_2^\Omega}\,) \wedge D(\,Da_1^\Omega +3i a_2^\Omega\,)  + A\wedge da_1^J \wedge da_1^J -A\wedge da_1^\Phi \wedge da_1^\Phi  \right. \\ [2mm]
\nonumber \!\!\!\!\!\!\!&& \displaystyle -\, \frac{1}{2}\,{\rm Re}\left[\big(\,Da_1^\Omega + 3ia_2^\Omega + f_2^\Omega\,\big)\wedge \left( b_2\wedge \overline{Dc^\Omega} + \overline{b^\Omega} (dc_2 -c_1dA) - b \leftrightarrow c \right)\right]\quad \\ [2mm]
\label{SCS}
\nonumber \!\!\!\!\!\!\!\!&& \displaystyle-\,\frac{1}{2}\big(\,da_1^J + f_2^J\,\big)\wedge\big[b_2\wedge Dc^J+b^J\left(dc_2-c_1\wedge dA\right)- \,\,b\leftrightarrow c\,\,\big]\\[2mm]
\nonumber \!\!\!\!\!\!\!\!&& 
 \displaystyle+\,\frac{1}{2}\big(\,da_1^\Phi + f_2^\Phi\,\big)\wedge\Big[\left(p\,c_2-q\,b_2\right)\wedge A+\big(b_2\wedge Dc^\Phi+b^\Phi\left(dc_2-c_1\wedge dA\right)- \,b\leftrightarrow c\,\big)\Big]\\[2mm]
\nonumber \!\!\!\!\!\!\!\!&&
 \displaystyle+\, \frac{1}{2}\,\,\big[ p(c_2+c_1\wedge A) - q(b_2 + b_1\wedge A)\big]\wedge \big[ c^\Phi d(b_2+b_1\wedge A)  - b^\Phi d(c_2 + c_1\wedge A) \big]\\ [2mm] 
\nonumber \!\!\!\!\!\!\!\!&&
 \displaystyle+\,\frac{1}{2}\left(Da+f_1\right)\wedge\left[b_2\wedge dc_1-b_1\wedge\left(dc_2-c_1\wedge dA\right)-\,\,b\leftrightarrow c\,\,\right]\\ [2mm]
 \!\!\!\!\!\!\!\!&& -\,\frac{1}{2}\,(k+f_0)\left[b_2\wedge \left(dc_2-c_1\wedge dA\right)-\,\,b\leftrightarrow c\,\,\right]\bigg\}.
\end{eqnarray}
%
Finally, the scalar potential $\mathcal V$ is
\begin{eqnarray}
S_{\rm pot}\!\!&\equiv&\!\!  \displaystyle\frac{1}{2\kappa_{5}^2}\int \big(-2\mathcal V\big) *1 \nonumber\\[2mm]
&=& \!\! \frac{1}{2\kappa_5^2}\,\int\,\Big\{e^{-\frac{8}{3}u-\frac{2}{3}v}\,R_{T^{1,1}}\,-\,2\,e^{-\frac{32}{3}u-\frac{8}{3}v}\,f_0^2\nonumber\\[2mm]
&&\qquad \;\;\, -\, e^{-\frac{20}{3}u-\frac{8}{3}v-\phi}\Big[{\rm Re}\left(-e^{-2i\theta}\shsh t\,(h_0^\Omega)^2 + 2p\, i e^{-i\theta}\,\sh(2t)\:\ch{(2w)}h_0^\Omega\right) \nonumber\\[2mm] 
&& \qquad\qquad\qquad\qquad\quad +\,\chch t \,|h_0^\Omega|^2\, +\, p^2\big(\chch t \,\ch{(4w)} + \shsh t\big)\Big]\nonumber \\ [2mm]
&& \qquad \;\;\, -\, e^{-\frac{20}{3}u-\frac{8}{3}v+\phi}\Big[\; h \,\to\, g\,,\quad p\,\to \,(q-p \,C_0)\; \Big]\Big\}*1\,, \label{eq:ScalarPot}
\end{eqnarray}
where $R_{T^{1,1}}$ was given in (\ref{R_T11}). 

We remark that by setting to zero the fields $a_1^\Phi,\,w,\, t,\, \theta,\, b^\Phi,\, c^\Phi$ and the 3-form flux parameters $p,\,q$, this 5-dimensional action reduces to the one given in \cite{IIBonSE}, describing a universal consistent truncation on squashed Sasaki--Einstein manifolds. We thus see that the consistent truncation to SU(2)$\times$SU(2) invariant modes presented here is an extension of the one based on general Sasaki--Einstein structures. In \cite{IIBonSE}, the latter was identified as a 5-dimensional gauged $\mathcal N=4$ supergravity with two vector multiplets. In the next section we prove that the present truncation also matches the structure of gauged $\mathcal N=4$ supergravity, with one more vector multiplet accommodating the extra fields included in the present $T^{1,1}$ case.

Before passing to the $\mathcal N=4$ supergravity analysis, we also remark that another subsector of our truncation ansatz coincides with the Papadopoulos--Tseytlin ansatz \cite{PapadopoulosTseytlin}. The precise dictionary is given in appendix \ref{RecoveringPT}. It follows that our supersymmetric consistent truncation includes the non-supersymmetric, constrained consistent truncation derived in \cite{BergHaackMueck}.

\section{Compatibility with $\mathcal N=4$ supergravity}\label{gaugedsugra}

The model coming from the left-invariant reduction of type IIB supergravity on $T^{1,1}$ presented in the last section matches the structure of gauged $\mathcal N=4$ supergravity in 5 dimensions. In the general case of an arbitrary manifold admitting a Sasaki--Einstein structure, the reasons to expect the reduction to preserve $\mathcal N=4$ supersymmetry were presented in \cite{IIBonSE}, where the procedure to match with the general formulation of  \cite{DHZ, SchonWeidner} was also detailed. In the case at hand, we have a new vector $a_1^\Phi$ coming from the expansion of $C_4$ in the cohomologically non-trivial three-form $\Phi\wedge\eta$, that in addition to the five new scalars fill the bosonic content of an $\mathcal N=4$ vector multiplet
\begin{equation}
\{a_1^\Phi, w,  t, \theta, b^\Phi, c^\Phi\}\,.
\end{equation}
Multiplets of this kind are sometimes dubbed ``Betti multiplets'' \cite{Dauria}, the reason being that they arise from an expansion in non-trivial cohomology representatives, and therefore their presence in the Kaluza--Klein spectrum relies on the non-vanishing Betti numbers of the internal manifold.

Since we have several fluxes turned on, the theory will be gauged. Taking also into account the two vector multiplets present in any squashed Sasaki--Einstein reduction, we end up with a  gauged $\mathcal N=4$ supergravity coupled to three vector multiplets, as we proceed to show.

\subsection{Identification of the $\mathcal N=4$ fields via the ungauged theory}

In order to identify the scalar and vector fields coming from the dimensional reduction with the ones of 5-dimensional $\mathcal N=4$ supergravity, we neglect all the interactions due to the flux parameters $k,p,q$, as well as those coming from the non-closure of the forms $\eta,J,\Omega$, and go therefore in the limit in which the theory is ungauged. After having properly identified the fields, in the next subsection we will analyse the effect of the gauging.

The scalar manifold has to be the 16-dimensional space
\begin{equation}\label{eq:ScalarManifold}
\mathcal M_{\rm scal} \,=\, {\rm SO}(1,1)\times \frac{{\rm SO}(5,3)}{{\rm SO}(5)\times {\rm SO}(3)}\,.
\end{equation}
Since SO(5)$\times$SO(3) is the maximal compact subgroup of SO$(5,3)$, we can parameterize the coset by exponentiating the solvable Lie subalgebra of $\mathfrak{so}$(5,3), which is spanned by a basis of non-compact generators (see e.g. \cite{Solvable}). For the generators of $\mathfrak{so}$(5,3) in the fundamental representation we take
\begin{equation}
(t_{MN})_P{}^Q \,=\, \delta^Q_{[M}\eta_{N]P}
\end{equation} 
where $\eta={\rm diag}\{-----+++\}$, and the indices run from $1$ to $8$. A basis for the solvable subalgebra of $\mathfrak{so}$(5,3) is given by the non-compact Cartan generators
\begin{equation}
C_1=t_{16}\,,\quad\quad\quad C_2=t_{27}\,,\quad\quad\quad C_3=t_{58}\,,
\end{equation}
together with
\begin{eqnarray}
G_1 \!\!\!&=&\!\!\! \txt{\frac{1}{2}}\left (t_{17} - t_{26} -t_{67} -t_{12}\right)\,, \quad\quad G_2 =\txt{\frac{1}{2}}\left(t_{17} +t_{26} - t_{67} + t_{12}\right)\,,\nonumber\\[2mm]
G_3 \!\!\!&=&\!\!\! \txt{\frac{1}{\sqrt 2}}\left(t_{36} + t_{13}\right)\,,\qquad\qquad\qquad G_{4}=\txt{\frac{1}{\sqrt 2}}\left(t_{37} +t_{23}\right)\,,\nonumber\\[2mm]
G_{5} \!\!\!&=&\!\!\! \txt{\frac{1}{\sqrt{2}}}\left(t_{46} +t_{14}\right)\,,\qquad\qquad\qquad G_{6}=\txt{\frac{1}{\sqrt{2}}}\left(t_{47} +t_{24}\right)\,,\nonumber\\ [2mm]
G_{7} \!\!\!&=&\!\!\! \txt{\frac{1}{\sqrt{2}}}\left(t_{56} +t_{15}\right)\,,\qquad\qquad\qquad G_{8}=\txt{\frac{1}{\sqrt{2}}}\left(t_{57} +t_{25}\right)\,, \nonumber\\ [2mm]
G_9 \!\!\!&=&\!\!\! \txt{\frac{1}{\sqrt 2}}\left(t_{18} - t_{68}\right)\,, \qquad\qquad\quad\;\; G_{10}=\txt{\frac{1}{\sqrt 2}}\left(t_{28} - t_{78}\right)\,,\nonumber\\[2mm]
G_{11} \!\!\!&=&\!\!\! \txt{\frac{1}{\sqrt{2}}}\left(t_{38} +t_{53}\right)\,,\qquad\qquad\quad\;\; G_{12}=\txt{\frac{1}{\sqrt{2}}}\left(t_{48} +t_{54}\right)\,,
\end{eqnarray}
which are closely related to the nilpotent, positive-root generators.\footnote{The positive-root generators are obtained defining the combinations $\frac{1}{\sqrt 2}(G_9\pm G_7)$ and $\frac{1}{\sqrt 2}(G_{10}\pm G_8)$.} Then we introduce coordinates $\{\phi_1,\phi_2,\phi_3,x_1,x_2,\ldots,x_{12}\}$ and parameterize the $\frac{{\rm SO}(5,3)}{{\rm SO}(5)\times {\rm SO}(3)}$ coset representative $L$ as
\begin{equation}
L \,=\, \left(\prod_{i=0}^{7} e^{x_{10-i} G_{10-i}}\right)
e^{x_{12}G_{12}}e^{x_{11}G_{11}}e^{x_2 G_2}e^{x_1 G_1}e^{\phi_3 C_3}e^{\phi_2 C_2}e^{\phi_1 C_1}\,.
\end{equation}
Our specific choice of generators and the order of the exponentials in $L$ is mainly dictated by computational convenience and ease of comparison with the supergravity fields.
 
Introducing the symmetric matrix
\begin{equation}
M_{MN}=\left(LL^T\right)_{MN},
\end{equation}
and its inverse $M^{MN}$, the metric on the scalar manifold (\ref{eq:ScalarManifold}) takes the form \cite{SchonWeidner}
\begin{equation}
-\frac{1}{2}ds^2\left(\mathcal{M}_{\rm scal}\right)\;=\; -\frac{3}{2}\Sigma^{-2}d\Sigma\otimes d\Sigma\,+\,\frac{1}{16}dM_{MN}\otimes dM^{MN}\,,
\end{equation}
$\Sigma$ being the scalar parameterizing the SO$(1,1)$ factor. Evaluating this expression in terms of the representative $L$ above, we recover precisely the scalar kinetic terms (\ref{scalkinterms}), provided we identify the Cartan coordinates as
\begin{eqnarray}
\phi_1&=&4u-\phi\nonumber\\
\phi_2&=&4u+\phi\nonumber \\
\phi_3&=& -4w -2 \log(\ch\, t)\,,
\end{eqnarray}
the $x$-coordinates as
\begin{eqnarray}
\nonumber x_1&=& 4a+ 2b^J c^J -2 b^\Phi c^\Phi +2{\rm Re}( b^\Omega\overline{c^\Omega})\,,\\[2mm]
\nonumber x_2 &=& 2C_0\\[2mm]
\nonumber \{x_3,x_5,x_7,x_9\} &=&2\sqrt{2} \left\{{\rm Re}(c^\Omega),{\rm Im}(c^\Omega),c^J,c^\Phi\right\}\,,\\ [2mm]
\nonumber \{x_4,x_6,x_8,x_{10}\} &=& 2\sqrt{2}\left\{{\rm Re}(b^\Omega),{\rm Im}(b^\Omega),b^J,b^\Phi\right\}\\[2mm]
\nonumber x_{11} &=& 2 \sqrt{2}\,e^{-2w}\tanh t \,\sin\theta\,,\\ [2mm] 
x_{12} &=& -2 \sqrt{2}\,e^{-2w}\tanh t \,\cos\theta\,,
\end{eqnarray}
and the SO$(1,1)$ scalar as
\begin{equation}
\Sigma=e^{-\frac{2}{3}(u+v)}\,.
\end{equation}

A similar philosophy can be adopted to identify the $\mathcal N=4$ supergravity vectors. The general form of the vector kinetic terms in the ungauged theory is 
\begin{equation}\label{eq:kinvectorsSW}
S_{\rm kin,vect} \;=\; -\frac{1}{2\kappa_5^2}\int  \left[\, \Sigma^{-4} \,(dA^0)^2 + \Sigma^2 \,M_{MN} dA^M\lrcorner \,dA^N   \,\right]*1,
\end{equation}
where the 1-forms are separated into a singlet $A^0$ and a fundamental $A^M$ representation of SO$(5,3)$. The 1-forms $A^0,A^1,\ldots,A^5$ belong to the $\mathcal N=4$ gravitational multiplet, while each of the remaining three enters in a vector multiplet. This field content is reproduced by our truncation by noticing that in the ungauged case the 2-form $a_2^\Omega$ drops from the action in favour of  $a_1^\Omega$, and moreover the 2-forms $b_2$ and $c_2$ can be dualized to 1-forms, which we call respectively $\widehat b_1$ and $\widehat c_1$. The explicit expression for the vector kinetic terms that we found after this dualization is long and we do not report it here. It is sufficient to tell that it matches the general $\mathcal N=4$ formula (\ref{eq:kinvectorsSW}) if we identify 
\begin{eqnarray}
	A \!\!&=&\!\! \sqrt2\, A^0\,, \,\,\,\qquad\quad b_1 = -A^1-A^6,\qquad \;\;\,
 c_1 = A^2+A^7,\qquad 
 \tilde a_1^J = A^5\,,\nonumber  \\[2mm]
 \widehat b_1 \!\!&=&\!\! -A^1+A^6,\qquad
 \widehat c_1 = A^2-A^7,\qquad
 {\rm Re} \,\tilde a_1^\Omega = A^3,\qquad  \,\;{\rm Im}	\, \tilde a_1^\Omega = A^4\,, \nonumber \\ [2mm] 
\tilde a_1^\Phi \!\!&=&\!\! A^8 \,,
\label{eq:IdentifVectors}
\end{eqnarray}
where the tildes denote the field redefinition given in (\ref{eq:FieldRedef}), needed to have proper abelian transformations.
The first two lines of (\ref{eq:IdentifVectors}) are as in \cite{IIBonSE}, while the further 1-form considered in this paper is identified with the new field $A^8$.

\subsection{The gauging and the embedding tensor}\label{EmbTensor}

Having fixed the frame of the ungauged theory, all the information needed to specify the gauging is encoded in the embedding tensor.\footnote{For a review of the embedding tensor formalism in connection with compactifications we refer to~\cite{SamtlebenLectures}. } This is the object which embeds the gauge group into the global duality group. At the same time, it determines the various couplings in the lagrangian arising from the gauging, including the scalar potential, as well as the fermionic shifts appearing in the supersymmetry transformations. It has components $f_{MNP}=f_{[MNP]}$ and $\xi_{MN}=\xi_{[MN]}$, that appear in the gauge-covariant derivative of the scalars as follows \cite{SchonWeidner}
\begin{equation}\label{CovDer}
DM_{MN}=dM_{MN}+2A^Pf_{P(M}{}^QM_{N)Q}+2A^0\xi_{(M}{}^QM_{N)Q}\,.
\end{equation}

To determine the non-vanishing components of the embedding tensor in our dimensional reduction, we match this general expression with the scalar covariant derivatives obtained in section \ref{DimRed}. To do this, we use the identifications for the scalars and for the vectors found above. The result is:
\begin{eqnarray}
\nonumber f_{125} \!\!&=&\!\! f_{256} = f_{567} = - f_{157} = -2\,,\\ [2mm]
\nonumber \xi_{34}\!\!&=&\!\! -3\sqrt 2\,,\\ [2mm]
\nonumber \xi_{12} \!\!&=&\!\! \xi_{17}=-\xi_{26}= \xi_{67} = -\sqrt 2\, k\,,\\ [2mm]
\label{eq:OurEmbTensor}
\nonumber \xi_{28} \!\!&=&\!\! \xi_{78}= -\sqrt 2\, p \,,\\ [2mm] 
\xi_{18} \!\!&=&\!\! \xi_{68}= -\sqrt 2 \,q\,.
\end{eqnarray}
The first three lines are precisely the same as in \cite{IIBonSE}; as already noticed there, the higher-dimensional origin of $f_{MNP}$ resides in the geometric flux associated with the non-closure of $\eta$, namely \hbox{$d\eta = 2J$}, while $\xi_{34}$ arises from the geometric flux $d\Omega = 3i\,\eta\wedge \Omega$, and the $\xi$'s in the third line come from the RR 5-form flux $k$. The components in the last two lines were zero in \cite{IIBonSE}, and arise in the present context  from the introduction of NSNS and RR 3-form fluxes, parameterized by $p$ and $q$ respectively.

From the embedding tensor we can now deduce the gauge group. We denote $(t_\Lambda)_M{}^N$, with $\Lambda=0,1,2,3$, the generators of the gauge group, and $A^\Lambda= \{A, b_1,c_1,\tilde a_1^J\}$ the associated gauge fields, in such a way that the gauge covariant derivative acting on the scalars reads $D = d- A^\Lambda t_{\Lambda}$. By comparison with the derivatives (\ref{CovDer}) we find that
\begin{eqnarray}
\nonumber t_0 &=& -6\,t_{34}+ 4k\,G_1 + 2\sqrt 2\, (q \,G_{9} + p\, G_{10})\,,\\ [2mm]
t_1 &=& 4\sqrt 2\,G_8\,,\qquad t_2 = 4\sqrt 2\,G_7\,,\qquad t_3= 8\,G_1\,, \label{eq:GaugeGen}
\end{eqnarray}
the only non-trivial commutator being $[t_1,t_2]=-2t_3\,$. It follows that the gauge group is $G={\rm Heis}_3 \times {\rm U}(1)$, as it was the case in \cite{IIBonSE}. Concerning the additional 1-form we have, $\tilde a_1^\Phi$, we point out that all the SU(2)$\times$SU(2) invariant scalars are neutral under it, so it doesn't really participate in the $\mathcal N=4$ supergravity gauging procedure.

In appendix  \ref{GaugeTransf} we show how one arrives at the same conclusions about the structure of the gauge group by directly reducing the 10-dimensional reparameterization invariance together with the gauge symmetry of the type IIB forms.

The remaining 1-forms that appeared in the ungauged picture also transform under the gauge group generators, though in a non-adjoint representation. The gauging procedure forces them to be dualized to tensor fields \cite{DHZ}, which are identified with the \hbox{2-forms} $b_2,\,c_2,\,\tilde a_2^\Omega$ appearing in our truncated action. For more details about this we refer to the analysis done in \cite{IIBonSE}.

We conclude this section with two comments. First we remark that, using the formula given in eq.~(3.16) of ref.~\cite{SchonWeidner}, we checked that the scalar potential (\ref{eq:ScalarPot}) matches its general $\mathcal N=4$ supergravity expression following from the embedding tensor above. Second, we stress that, having the precise form of the embedding tensor, it is now straightforward to write down the fermionic supersymmetry variations for our $\mathcal N=4$ model, using the general formulae provided in \cite{SchonWeidner}. It would be very interesting to study the known conifold solutions from this perspective, and possibly search for new supersymmetric solutions. In particular, this formalism should be useful for finding a superpotential providing first-order equations for the backgrounds. We hope to report on these issues in the near future.

\section{A new family of AdS$_5$ solutions}\label{NewFamilyAdS}

In the following we search for AdS$_5$ solutions of type IIB supergravity on $T^{1,1}$ by extremizing the 5-dimensional scalar potential $\mathcal V$, given in (\ref{eq:ScalarPot}). The lifting of the solution is guaranteed by the consistency of our truncation. One can see that if the 3-form flux parameters $p$ and/or $q$ are non-vanishing, then there are no extrema at finite values of the fields (the variation with respect to $c^\Phi$ or $b^\Phi$ removes the RR 5-form contribution $f_0^2$ from $\mathcal V$). Thus we focus on $p=q=0$, leaving just the RR 5-form flux $k$ switched on. In this case a first extremum of $\mathcal V$ corresponds to the well-known supersymmetric solution of \cite{RomansNewIIBsol}, which for $k=2$ picks the Sasaki--Einstein metric (\ref{eq:SEpoint}). We will come back to this background in section~\ref{DualOperators}, where we will discuss the spectrum of the field fluctuations in the context of the gauge/gravity duality. No other supersymmetric AdS$_5$ solutions are expected \cite{SusyAdSIIB}.

Interestingly, we find a family of non-supersymmetric AdS$_5$ solutions, which for $k=2$ is given by
\begin{eqnarray}
\nonumber e^{8u} \!\!&=&\!\! e^{- 8v} \,=\, \frac{4}{9}\,\ch (2t)\;,\qquad\qquad \ch (2w) \,=\, \frac{\ch^{\frac{1}{2}}(2t)}{\ch \,t}\;,\\ [2mm]
\qquad b^\Omega  \!\!&=&\!\! e^{\frac{\phi}{2}+i\theta}\,\frac{\big[2-\ch(2t)\big]^{\frac{1}{2}}}{3^{\frac{1}{2}}\,\ch^{\frac{1}{4}} (2t)}\;, \quad\qquad c^\Omega \,=\, \left[ C_0 + i\,e^{-\phi}\ch^{\frac{1}{2}}(2t) \right] b^\Omega,\label{eq:FamilySols}
\end{eqnarray}
with arbitrary $C_0,\,\phi,\,\theta$, and with $t$ being bounded between 0 and $\ch(2t) = 2\;\Leftrightarrow t\simeq 0.66\;$ (the remaining scalars $b^J,c^J,b^\Phi,c^\Phi,a$ don't appear in $\mathcal V|_{p=q=0}$, so their value is obviously free).\footnote{We still have an extremum if in the above solution we replace $e^{\phi}$ with $e^{\phi}\ch (2t)$ and $\theta$ with $\theta + \frac{3}{2}\pi$.} The AdS cosmological constant, corresponding to the value taken by the potential at the extremum, is $\Lambda = -\frac{27}{4}$.

We can regard (\ref{eq:FamilySols}) as a family of solutions described by the $T^{1,1}$ metric parameter $t$, the other moduli entering in a rather trivial way. At the lowest value of $t$ we have
\begin{equation}
t=w=0\,,\qquad  e^{4u}=e^{-4v}=\frac{2}{3}\,,\qquad  b^\Omega= \frac{e^{\frac{\phi}{2} +i \theta}}{3^{\frac{1}{2}}}\,\,,\qquad c^\Omega=  (C_0  + i\,e^{-\phi}) \, b^\Omega  \,,
\end{equation}
where here $\theta$ just represents an arbitrary phase. The corresponding solution of type IIB supergravity was found long ago by Romans \cite{RomansNewIIBsol}, applying a construction previously employed by Pope and Warner in the context of 11-dimensional supergravity \cite{PopeWarner}. This is based on building the compact 5-dimensional manifold as a U(1) fibration over any K\"ahler--Einstein base, and works in particular for $T^{1,1}$, which is a U(1) fibration over $\mathbb{CP}^1\times \mathbb{CP}^1$. The resulting metric on the 5-dimensional manifold is non-Einstein. 

To the best of our knowledge, for non-zero $t$ the solutions (\ref{eq:FamilySols}) are new. It is especially interesting to look at the upper extremum of the range of $t\,$:
\begin{equation}\label{eq:NewFreundRubinSol}
\ch (2t)\,=\, 2\,,\;\qquad e^{4w}\,=\,3\,,\;\qquad e^{8u}\,=\,e^{-8v}\,=\, \frac{8}{9}\,,\qquad\;  b^\Omega = c^\Omega = 0\;.
\end{equation}
Since at this point the type IIB 3-form field strengths $H$ and $F_3$ vanish, we have an AdS$_5$ solution of the Freund--Rubin type, supported by just the 5-form flux \cite{FreundRubin}. Recalling the formulae in section \ref{T11MetricCurvature}, one can check that the $T^{1,1}$ metric defined by (\ref{eq:NewFreundRubinSol}) is indeed Einstein, as expected for a Freund--Rubin solution. This Einstein metric was known in the mathematical literature \cite{Alekseevsky}, but to our knowledge it had not been considered previously in the context of string theory compactifications. 

We have thus found a family of non-supersymmetric AdS$_5$ solutions of type IIB supergravity interpolating between the Romans--Pope--Warner background and a Freund--Rubin solution employing an Einstein metric on $T^{1,1}$ not related to the usual Sasaki--Einstein one.

That the solution (\ref{eq:FamilySols}) is not supersymmetric can also be deduced by the fact that there is no massless graviphoton on it: indeed, for all values of $t$ the background is charged under the U(1) gauge field $A$, which then acquires a mass via the Higgs mechanism.

Since we are dealing with non-supersymmetric solutions, a basic issue is the one of stability. In particular, this is crucial for the dual conformal field theory to be well-defined. If any affirmative proof of stability can only be given by working at the untruncated level, below we reach some negative conclusions by studying the mass spectrum of the SU(2)$\times$SU(2) invariant modes entering in our scalar potential, and showing that for~$t$ beyond a certain value the solutions above are unstable.

For the Romans--Pope--Warner solution we enhance the analysis already done in \cite{IIBonSE} to the full set of SU(2)$\times$SU(2) invariant modes, by including the fluctuations of $b^\Phi,\,c^\Phi,w,\,\,t$ and $\theta$. Looking at the mass eigenstates, we find that these fluctuations don't mix with the other ones, and are all massless. Recalling that the masses squared of the modes previously considered in \cite{IIBonSE} were also all non-negative, we conclude that there is no hint of instability of the Romans--Pope--Warner solution among the left-invariant modes on $T^{1,1}$. Of course, it is possible that some of the non-left-invariant modes, to which we don't have access here, turn out to be unstable. This expectation is supported by the fact that the analogous solution on the 5-sphere $S^5$ is known to be unstable \cite{GPPZ2}.

\begin{figure}[!h]
	\centering
\includegraphics[scale=.37]{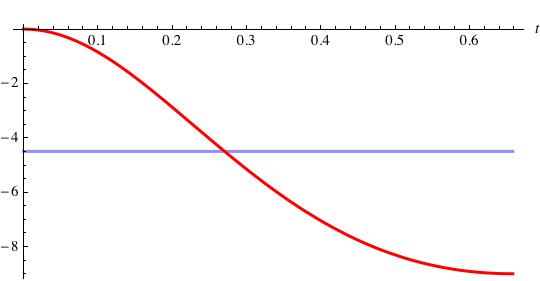} 
\caption{\small Plot of the lowest mass$^2$ eigenvalue as a function of the $T^{1,1}$ metric parameter $t$. The horizontal line is the Breitenlohner--Freedman bound.}
\label{MinMassPlot}
\end{figure}

The situation is worse for solution (\ref{eq:NewFreundRubinSol}). In this case, the non-vanishing masses are
\begin{equation}
m^2 \,=\, \{45,\,36,{\textstyle{\frac{63}{2}}},\,{\textstyle{\frac{63}{2}}},\,9,\,-9\}\,,
\end{equation} 
with the negative value violating the Breitenlohner--Freedman bound \cite{BFbound}, which in 5 dimensions reads $m^2 \geq -\frac{2}{3}|\Lambda|\;$ \cite{MezincescuTownsend}. Hence this background is unstable. Concerning the interpolating solutions, we find positive masses at any value of $t$, except for one mode, whose behaviour is given in figure~\ref{MinMassPlot}: it is massless at the Romans--Pope--Warner  solution ($t=0$), then its mass squared becomes progressively more negative while $t$ increases, violating the Breitenlohner--Freedman stability bound at $t \simeq 0.27$.

\section{The dual operators}\label{DualOperators}

In this section we put our consistent truncation in the perspective of the gauge/gravity duality: we discuss the mass spectrum of the field fluctuations about the supersymmetric AdS$_5\times T^{1,1}$ vacuum, and identify their dual operators. 

In \cite{KlebanovWitten}, the field theory dual to type IIB string theory on the supersymmetric AdS$_5\times T^{1,1}$ background with $N$ units of $F_5$ flux was shown to be an $\mathcal N=1$ superconformal theory with gauge group ${\rm SU}(N)\times{\rm SU}(N)$ and global symmetry group given by the product of a flavour ${\rm SU}(2)\times{\rm SU}(2)$ with the ${\rm U}(1)$ R-symmetry. The global ${\rm SU}(2)\times{\rm SU}(2)$ is identified with the one naturally acting from the left on the $T^{1,1}$ coset, while the R-symmetry arises from the right-isometry corresponding to the reparameterization of the U(1) fibre. The degrees of freedom in the theory consist of the gauge superfields $W_{1\,\alpha},\,W_{2\,\alpha}$ ($\alpha$ is a spinorial index) for the two factors of the gauge group, together with chiral superfields $A_i,\,B_j\,,\;i,j=1,2$, with $A$ transforming in the $( {\rm N}, \overline {\rm N})$ representation of the gauge group and being a doublet under the first global SU(2), and $B$ transforming in the $( \overline {\rm N}, {\rm N})$ and being a doublet under the second SU(2). 

We can immediately observe that, since the type IIB supergravity modes preserved by our truncation are precisely the ${\rm SU}(2)\times{\rm SU}(2)$ invariant ones, their dual operators have to be flavour singlets. Also taking into account the fermionic nature of $W_{1\,\alpha},\,W_{2\,\alpha}$, there is only a finite number of such operators. To identify which ones among these are dual to the supergravity fields in our truncation, we use the results of \cite{T11spectrum}, where the full spectrum of type IIB supergravity on the supersymmetric AdS$_5\times T^{1,1}$ background was derived, and the matching with the SCFT operators was established. 

We start noticing that on this background, the 5-dimensional $\mathcal N=4$ supergravity fields reorganize into massless and massive $\mathcal{N}=2$ multiplets. This is due to the spontaneous partial breaking of supersymmetry and gauge symmetry driven by the various St\"uckelberg couplings in the truncated action. In addition to the massless graviton, massive gravitino, massive vector and hyper multiplets constituting the universal sector of the Kaluza--Klein spectrum of type IIB supergravity on Sasaki--Einstein manifolds \cite{IIBonSE}, in the present $T^{1,1}$ case we have the $\mathcal N=4$ Betti vector multiplet $\{a_1^\Phi, w,b^\Phi, c^\Phi, t, \theta\}$. This splits in an $\mathcal N=2$ vector multiplet, including  the fluctuation of the scalar $w$ and the massless gauge field $a_1^\Phi$, and in a hypermultiplet, whose bosonic content is given by the fluctuations of the scalars $b^\Phi,c^\Phi,t$ and $\theta$. It is using this  $\mathcal N=2$ picture that one can establish a direct relation between the supergravity multiplets and the $\mathcal N=1$ SCFT superfields.

\renewcommand{\arraystretch}{1.2}
\begin{table}
\begin{center}
$	\begin{array}{rcccccccl}\hline
\mathcal N=2\: {\rm multiplet} && {\rm field\: fluctuations} && m^2 &\phantom{s}& \Delta & \phantom{s}& {\rm dual\; operators}  \\\hline
\rule{0pt}{3ex}
{\rm gravity} && \begin{array}{c} A - 2a_1^J \\ g_{\mu\nu} \end{array}&& \begin{array}{c} 0 \\ 0 \end{array} && \begin{array}{c} 3\\ 4 \end{array} &&  {\rm Tr}(W_{1\alpha}\overline W_{\!1 \dot\alpha} + W_{2\alpha}\overline W_{\!2 \dot\alpha} )+\ldots\\ \hline
{\rm universal \;hyper} && \begin{array}{c} b^\Omega - i \,c^\Omega \\ \phi\,,\;\;C_0 \end{array} && \begin{array}{c} -3 \\ 0 \end{array} && \begin{array}{c} 3 \\ 4 \end{array} && {\rm Tr} (W_{\!1}^2 + W_{\!2}^2) +\ldots   \\ \hline
{\rm Betti \;vector} && \begin{array}{c} w \\ a_1^{\Phi} \end{array} && \begin{array}{c} -4 \\ 0\end{array} && \begin{array}{c} 2\\ 3 \end{array} && {\rm Tr}\,A e^{V_2} \overline A e^{-V_1} -{\rm Tr}\, B e^{V_1}\overline B e^{-V_2} \\ \hline
{\rm Betti \;hyper} &&	\begin{array}{c}	 t \,e^{i\theta}\\  b^\Phi,\;\;c^\Phi \end{array} && \begin{array}{c} -3 \\   0   \end{array} && \begin{array}{c} 3 \\  4   \end{array}  &&  {\rm Tr} (W_{\!1}^2 - W_{\!2}^2)  \\  \hline
		&& b_1,\;c_1 && 8 && 5 && \\
{\rm massive\; gravitino} && a_2^\Omega && 9 && 5 &&  {\rm Tr}(W_{\!1}^2\overline{W}_{\!1\dot \alpha} + W_{\!2}^2\overline{W}_{\!2\dot \alpha}) + \ldots \\ 
&&		b_2,\;c_2 && 16 && 6  && \\ \hline
{\rm massive \;vector} && \begin{array}{c} u - v\\ A + a_1^J\\ b^\Omega + i \,c^\Omega \\ 4u + v \end{array}&&\begin{array}{c} 12\\ 24 \\ 21\\ 32  \end{array}  &&\begin{array}{c}  6\\   7 \\  7\\  8 \end{array} &&  {\rm Tr}(W_{\!1}^2\overline W{}_{\!1}^2 + W_{\!2}^2\overline W{}_{\!2}^2) + \ldots \\ \hline
\end{array}$
\caption{Mass eigenstates of the full SU(2)$\times$SU(2) invariant bosonic sector of type IIB supergravity on the supersymmetric AdS$_5\times T^{1,1}$ background, and their dual superfield operators. The mass eigenvalues are evaluated choosing the RR flux $k=2$, which yields a unit AdS radius; the corresponding mass eigenstates are given for zero vevs of the moduli $C_0,\phi$.  The massive vectors $b_1,\,c_1,\, A + a_1^J$ have eaten their respective axions $b^J,\,c^J,\,a$.} \label{GaugeGravityTable}
\end{center}
\end{table}

The results following from the diagonalization of the mass matrix are summarized in table~\ref{GaugeGravityTable}, where we provide the mass eigenstates entering in each $\mathcal N=2$ multiplet, together with their mass eigenvalues, the conformal dimension $\Delta$ of the corresponding dual operators, and the dual superfields accommodating the single operators. 
As one can see, not all the possible superfields being ${\rm SU}(2)\times{\rm SU}(2)$ singlets appear in the table: some of these operators indeed develop large anomalous dimensions in the large $N$ limit, meaning that they are really stringy states which cannot be captured by the supergravity approximation. In particular, this is the case for the combinations of the gauge superfields $W_{\!1\alpha},\,W_{\!2\alpha}$ involving a minus sign \cite{BaumannEtAlD3potentials, CDSWchiralrings}, with the remarkable exception of ${\rm Tr} (W_{\!1}^2 - W_{\!2}^2)$, which has instead protected conformal dimension.

The inclusion of the Betti sector in addition to the degrees of freedom already studied in \cite{IIBonSE} allows us to take into account not only the sum of the complexified couplings of the two SCFT gauge groups, which is dual to the axio-dilaton $C_0+ie^{-\phi}$, but also the respective difference, which is dual to the scalars $b^\Phi$ and $c^\Phi$ \cite{KlebanovWitten}. These latter modes play a crucial role in the Klebanov--Strassler solution, where they are responsible for the running of the gauge coupling and for the chiral anomaly \cite{KlebanovStrassler, KlebanovOuyangWitten}. The operator dual to the Betti vector multiplet is the baryonic current multiplet\footnote{In table~\ref{GaugeGravityTable}, the $V_1,\,V_2$ appearing in the expression for the baryonic current multiplet are the vector superfields of which $W_{\!1\alpha}$ and $W_{\!2\alpha}$ are field-strengths \cite{T11spectrum}.} \cite{KlebanovWitten2}, where the ``baryon number'' symmetry of the field theory, acting on the chiral superfields as $A_i \to e^{i\zeta} A_i$ and $B_j \to e^{-i\zeta} B_j$, is dually identified with the abelian gauge symmetry of $\tilde a_1^\Phi$.

On the field theory side, the consistency of our truncation translates into the fact that the set of operators appearing in table~\ref{GaugeGravityTable} has to be closed under the operator product expansion (at least in the large $N$ limit). It would be interesting to give an explicit proof of this (see \cite{SkenderisTaylorTsimpis} for the relation between consistency of the truncation and closure of the OPEs in a closely related context).

\section{$\mathcal{N}=2$ subtruncations and backreacting D7-branes}\label{N=2subtrunc}

In this section we explore further consistent truncations preserving $\mathcal N=2$ supersymmetry. As we will show, in this more restricted setup it is possible to describe the deformation of the type IIB background by the addition of smeared D7-brane sources.

\subsection{Truncating to $\mathcal{N}=2$}

Switching off the full Betti vector multiplet is an obvious supersymmetric consistent truncation preserving $\mathcal N=4$, with the result being supported by any manifold admitting a Sasaki--Einstein structure \cite{IIBonSE}, not just $T^{1,1}$. Besides that, the model presented above admits some interesting consistent subtruncations preserving just $\mathcal N=2$ supersymmetry in 5 dimensions.\footnote{For the general structure of gauged $\mathcal N=2$ supergravity in 5 dimensions we refer to \cite{CeresoleDallAgataGaugedN=2}.} To understand which degrees of freedom can be truncated, it is useful to recall the discussion in the previous section, where we showed how for small fluctuations about the supersymmetric vacuum the 5-dimensional fields organize in massless and massive $\mathcal N=2$ multiplets. This $\mathcal N=2$ structure is expected to extend at the non-linear level, and should provide an alternative picture to our $\mathcal N=4$ gauged supergravity description, though probably less standard because of the presence of the massive multiplets.

The first step in halving the amount of supersymmetry is to truncate the $\mathcal N=2$ massive gravitino multiplet: as far as the bosonic fields are concerned, this means to take
\begin{equation}
b_2=c_2= a_2^\Omega = b_1=c_1= b^J=c^J=0\,,
\end{equation}
requiring at the same time that the corresponding equations of motion are exactly satisfied. One realizes that this consistency requirement can be fulfilled in two different ways: from the $\mathcal{N}=4$ Betti vector multiplet $\{a_1^\Phi, w,b^\Phi, c^\Phi, t, \theta\}$, viewed as the union of an $\mathcal{N}=2$ vector multiplet $\{a^\Phi_1, w\}$ and a hypermultiplet $\{b^\Phi, c^\Phi, t, \theta\}$, one can either keep the Betti hyper while truncating away the vector multiplet, or, vice versa, retain the Betti vector while switching off the hypermultiplet as well as the 3-form fluxes $p$ and $q$. 

Hence we end up with two different $\mathcal{N}=2$ supergravity models. The first one contains, apart from the gravity multiplet, three hypers plus a vector. The complete set of fields is
\begin{eqnarray}
&&\{g_{\mu\nu}, A, a_1^J, u+v\}\quad \;\qquad\qquad \textrm{gravity + vector} \nonumber \\ [1mm]
&&\{b^\Omega, b^\Phi, c^\Omega, c^\Phi, a, \phi, C_0, u, t, \theta\} \qquad\qquad \textrm{3 hypers,} 
\end{eqnarray}
where, for $k=2$, the combination $A-2a_1^J$ plays the role of the $\mathcal N=2$ graviphoton. We are splitting the massive vector multiplet appearing in the previous section in a standard vector multiplet $\{ a_1^J + A, u + v \}$ and a hypermultiplet. The latter contains the scalar $a$, which is St\"uckelberg-coupled to the vector and therefore makes it massive. Both the gravity and the vector multiplets come from the $\mathcal N=4$ gravity multiplet. The action which follows from truncating in this way the model given in section \ref{eq:5dmodel} has kinetic terms
\begin{eqnarray}
S_{\rm kin}\!\!\!&=&\!\!\!\frac{1}{2\kappa_5^2}\int \Big\{R-\frac{28}{3}du^2-\frac{4}{3}dv^2- \frac{8}{3}du \lrcorner dv - dt^2 - \shsh t \,(d\theta - 3A)^2  \nonumber\\[2mm]
&&\qquad   -\, e^{-4u-\phi}\Big[\ch(2t)\,(h_1^\Phi)^2+\,\chch t \,|h_1^\Omega|^2- \shsh t\, {\rm Re}\left(e^{-2i\theta} (h_1^\Omega)^2\right)\nonumber\\[2mm]
&& \qquad \qquad \qquad   +\,2 \,\sh(2t)\,h_1^\Phi\lrcorner\, {\rm Re}\big(i \,e^{-i\theta} h_1^\Omega \big) \Big]- e^{-4u+\phi}\Big[\,h\,\rightarrow\,g\,\Big] \nonumber\\[2mm]
&& \qquad     -\, \frac{1}{2}d\phi^2 - \frac{1}{2}e^{2\phi}dC_0^2 - 2\,e^{-8u}f_1^2 -\frac{1}{2}e^{\frac{8}{3}u+\frac{8}{3}v}(dA)^2-e^{-\frac{4}{3}u-\frac{4}{3}v}(da_1^J)^2 \Big\}\!*\!1,\quad
\end{eqnarray}
where the quaternionic manifold describing the scalar $\sigma$-model is $\frac{{\rm SO}(4,3)}{{\rm SO}(4)\times {\rm SO}(3)}$. The topological interactions simplify considerably, and read
\begin{equation}
S_{\rm top}\, =\, \frac{1}{2\kappa_5^2}\,\int\,A\wedge da_1^J\wedge da_1^J\,,
\end{equation}
while the scalar potential is basically the same as the complete $\mathcal{N}=4$ one, with the variable $w$ (always appearing as the argument of a hyperbolic cosine) set to zero. We can keep all the flux parameters $k,p,q$. Notice that this provides an $\mathcal N=2$ supersymmetric completion to the subsector of the Papadopoulos--Tseytlin ansatz that gives the Klebanov--Strassler solution \cite{KlebanovStrassler}. On the other hand, it does not contain the Maldacena--Nu\~nez one \cite{MaldacenaNunez} because in that case both $w$ and $t$ are non-vanishing, meaning that to render the corresponding subsector of the Papadopoulos--Tseytlin ansatz supersymmetric one should go to the full $\mathcal{N}=4$ theory.

The second $\mathcal N=2$ model consists of the gravity multiplet, two hypermultiplets and two vector multiplets, with field content
\begin{eqnarray}
&&\{g_{\mu\nu}, A, a_1^J, u+v\}\qquad\qquad \textrm{gravity + vector}  \nonumber\\ [1mm]
&&\{a_1^\Phi, w\}\qquad\qquad\qquad\qquad \;\, \textrm{Betti vector}   \nonumber \\ [1mm]
&&\{b^\Omega, c^\Omega, a, \phi, C_0, u\} \qquad\qquad \textrm{2 hypers.}
\end{eqnarray}
We also require $p=q=0$, i.e. no 3-form flux. The kinetic terms this time take the form
\begin{eqnarray}
S_{\rm kin}\!\!&=&\!\!\frac{1}{2\kappa_5^2}\!\int \Big\{R-\frac{28}{3}du^2-\frac{4}{3}dv^2- \frac{8}{3}du \lrcorner dv - 4dw^2 -\, \frac{1}{2}d\phi^2 \,-\, \frac{1}{2}e^{2\phi}dC_0^2\nonumber\\[2mm]
&& \qquad \quad  \,  - e^{-4u-\phi}\,|h_1^\Omega|^2- e^{-4u+\phi}\,|g_1^\Omega|^2\,- 2\,e^{-8u}\,f_1^2 -\frac{1}{2}e^{\frac{8}{3}u+\frac{8}{3}v}\,(dA)^2\nonumber\\[2mm]
&& \qquad \quad   \, -e^{-\frac{4}{3}u-\frac{4}{3}v}\,\ch(4w)\Big[(da_1^J)^2 +(da_1^\Phi)^2-2\tanh(4w)da_1^J\,\lrcorner \,da_1^\Phi\Big] \Big\}*\!1,
\end{eqnarray}
while the topological term is 
\begin{equation}
S_{\rm top}\,=\,\frac{1}{2\kappa_5^2}\,\int\,\left(A\wedge da_1^J\wedge da_1^J-A\wedge da_1^\Phi\wedge da_1^\Phi \right) \,,
\end{equation}
and the potential reads
\begin{eqnarray}
S_{\rm pot}&=&\frac{1}{2\kappa_5^2}\,\int\Big\{24\,e^{-\frac{14}{3}u-\frac{2}{3}v}\,\,\ch(2w)-4\,e^{-\frac{20}{3}u+\frac{4}{3}v}\,\,\ch(4w)-2\,e^{-\frac{32}{3}u-\frac{8}{3}v}\,f_0^2\nonumber\\[2mm]
&&\qquad\quad\,-\,e^{-\frac{20}{3}u-\frac{8}{3}v}\Big[e^{-\phi}\,|h_0^\Omega|^2+e^{\phi}\,|g_0^\Omega|^2\Big]\Big\}*1\,.
\end{eqnarray}
In this case, the quaternionic manifold is $\frac{{\rm SO}(4,2)}{{\rm SO}(4)\times {\rm SO}(2)}\,$, while the scalar manifold of the vector multiplets is $\big[$SO$(1,1)\big]^2$. From this model one can further consistently truncate the universal hypermultiplet by setting
\begin{equation}
\tau \equiv C_0+i\,e^{-\phi}= {\rm const}\,,\quad\quad\quad\quad\quad c^\Omega=\overline{\tau}\,\,b^\Omega\,.
\end{equation}
This provides a simple supersymmetrization of the consistent truncation derived in \cite{HerzKlebPufuTesi}, where black 3-brane solutions charged under the gauge field dual to the baryonic current ($a_1^\Phi$ in our notation) were found.

\subsection{Adding D7-branes}

These truncations with reduced supersymmetry are also suitable for the introduction of fundamental matter in the dual field theory. This is achieved by considering $N_f$ D7-branes in the type IIB supergravity \cite{KarchKatz}.\footnote{We thank Aldo Cotrone for drawing our attention to this possibility.} Our truncation can accommodate the backreacting branes as long as they are smeared in the transverse directions. To be definite, we will consider the particular setup of \cite{Ramallo}, where the smearing is described precisely by the 2-form $J$, determining the distribution of the RR charge of the D7-branes through the Wess--Zumino term. As explained there, this additional term in the action drives a magnetic coupling between $F_1$ and the branes that alters correspondingly its Bianchi identity. The RR 1-form is no longer closed but verifies, in the normalization of \cite{Ramallo},
\begin{equation}
dF_1=-\,\frac{6N_f}{4\pi}\,J\,.
\end{equation}
It is easy to solve this equation using the left-invariant 1-form $\eta$, giving a charge to the RR scalar
\begin{equation}
F_1\,=\, dC_0+F_1^{\rm D7} \,=\, DC_0+\,n\,(\eta+A)\,,
\end{equation}
where the covariant derivative reads $DC_0=dC_0-\,n\,A$ and we have introduced a non-closed term $F_1^{\rm D7}=n\,\eta\,$ sourced by the brane. The charge of the axion is $n=-\,\frac{3N_f}{4\pi}$ in the normalization adopted. This forces us to change accordingly the solution to the Bianchi identity for the RR 3-form. In order to solve it globally, we can take for instance
\begin{equation}
F_3\,=\, dC_2+F_3^{\rm fl}-C_0\,H+F_1^{\rm D7}\wedge B\,,
\end{equation}
but, due to the non-closure of $F_1$, we also need to impose $J\wedge B=0$.
Of course, the latter condition is not true in the general $\mathcal{N}=4$ reduction due to the presence of $b_2$, $b_1$ and $b^J$. Nevertheless, notice that all these fields are part of the gravitino multiplet, so they are not included in the $\mathcal N=2$ truncations presented above, that consequently verify automatically the desired condition and hence are appropriate for supporting supersymmetric flavoured solutions. 

The ansatz for the metric and the forms written in \cite{Ramallo} is a generalization of the Klebanov--Strassler one and, along with the D7-branes, contains the following fields
\begin{equation}
\{b^\Phi, {\rm Im}\,b^\Omega, {\rm Re}\,c^\Omega, \phi, u, v, w, t\}
\end{equation}
plus the RR 3-form flux $q$ and the new parameter $n$. We observe that, beside $w$, all the fields in the ansatz are present in the $\mathcal N=2$ truncation containing the Betti hypermultiplet. Fortunately, in both solutions given in \cite{Ramallo}, namely the flavoured warped deformed conifold and the flavoured Klebanov--Tseytlin, the variable $w$ plays no role and can be switched off. 
We conclude that the $\mathcal N=2$ truncations discussed in the previous section can accomodate backreacting smeared D7-branes, and are therefore a convenient arena to deal with supersymmetric flavoured conifold solutions. In particular, the truncation containing the Betti hypermultiplet, suitably generalized as outlined above, provides the solutions found in \cite{Ramallo}. 

Concerning the new parameter introduced, $n$, it changes the direction of one of the isometries being gauged. Under reparameterization of the U(1) fibre of $T^{1,1}$, the RR scalar now shifts and, compared with the equations in appendix \ref{GaugeTransf}, triggers an additional term in the transformation of $c^\Phi\,$:
\begin{equation}
\delta C_0=\,n\,\omega\,,\qquad\qquad\qquad\delta c^\Phi=\,\,q\,\omega\,+n\,\omega\,b^\Phi\,,
\end{equation}
supplemented by the field redefinition $\tilde{c}^\Omega= c^\Omega-C_0b^\Omega$ to get a proper charged complex scalar. Nevertheless, since in this truncation we have just two vectors, $a_1^J$ and $A$, the gauge group remains the product of two abelian factors, one of them being the R-symmetry of the dual.

\section{Outlook}\label{conclusions}

In this paper we have presented a consistent truncation of type IIB supergravity on the $T^{1,1}$ coset manifold, leading to a 5-dimensional gauged $\mathcal N=4$ supergravity model with three vector multiplets. Our supersymmetric truncation incorporates the one based on the Papadopoulos--Tseytlin ansatz \cite{PapadopoulosTseytlin, BergHaackMueck}, and is therefore suitable for studying the dynamics associated with the various solutions contained there, like the Klebanov--Strassler \cite{KlebanovStrassler} and the Maldacena--Nu\~nez \cite{MaldacenaNunez} ones, as well as the interpolating solution of \cite{BaryonicBranchKS}. 

To perform the dimensional reduction we exploited the coset structure of $T^{1,1}$, which provides a simple identification of the SU(2)$\times$SU(2) invariant modes of type IIB supergravity. The truncation ansatz preserving all and only these left-invariant modes is guaranteed to be consistent.
Although we focused just on the bosonic sector, for which we established a precise matching with the general structure of gauged $\mathcal N=4$ supergravity, we expect the fermionic sector to work accordingly. It should not be too hard to check this by studying the dimensional reduction of the supersymmetry variations. 

There is a number of interesting problems that could be addressed by taking advantage of the five-dimensional setup. For instance, it would be nice to explicitly determine the superpotential generating the full set of equations for the supersymmetric solutions in \cite{BaryonicBranchKS, MaldacenaMartelli}, as well as to see if the non-supersymmetric charged black 3-brane solution of \cite{HerzKlebPufuTesi} admits a superpotential. Concerning the latter case, we observe that indeed the existence of the superpotential does not require the solution to be supersymmetric \cite{FakeSugra}, and that it would nevertheless ensure its stability against the SU(2)$\times$SU(2) invariant fluctuations. 

Similar supersymmetric consistent truncations based on a left-invariant ansatz can certainly be derived by considering 11-dimensional supergravity on the 7-dimensional coset spaces admitting a Sasaki--Einstein structure and having non-trivial topology, such as $M^{1,1,1}$ and $Q^{1,1,1}$ (see e.g. \cite{KKreview} for a review of AdS$_4$ solutions of 11-dimensional supergravity on coset spaces). In these cases, the inclusion of the cohomologically non-trivial forms allows to take into account an internal 4-form flux. The presence of the Betti multiplets enhances the consistent truncation to 4-dimensional gauged $\mathcal N=2$ supergravity established in \cite{GauntlettKimVarelaWaldram}, and should also incorporate the non-supersymmetric reduction for  charged membrane solutions of~\cite{Klebanov11d}. Moreover, it should be easy to verify if the new solutions we found in section~\ref{NewFamilyAdS}, in particular the Einstein metric with non-vanishing off-diagonal parameters, also exist in these 7-dimensional cases.

Finally, it would be interesting to prove that the consistent truncation presented in this paper can be adapted to a truncation of type IIB supergravity on the non-homogeneous $Y^{p,q}$ and $L^{p,q,r}$ Sasaki--Einstein manifolds \cite{Ypq, Lpqr}. Since the topology of these spaces is the same as for $T^{1,1}$, namely $S^2\times S^3$, one can add to the basis of expansion forms used in \cite{IIBonSE} a 2-form and a 3-form being cohomologically non-trivial, and see if the corresponding truncation retaining the Betti vector multiplet is consistent. As suggested by the results of \cite{IIBonSE, BuchelLiu}, the coset structure of the compact manifold might not be a strictly necessary ingredient for deriving the consistent truncation. In general, it seems reasonable to expect that a truncation including all the modes whose dual field theory operators are flavour singlets be consistent.

\section*{Acknowledgments}

\noindent We are grateful to Gianguido Dall'Agata for many illuminating discussions and valuable comments. We also thank Irene Amado, Sergio Benvenuti, Aldo Cotrone, Nick Halmagyi and Dario Martelli for useful discussions. DC is supported by the Fondazione Cariparo Excellence Grant {\it String-derived supergravities with branes and fluxes and their phenomenological implications}.

\appendix

\section{Reduction formulae}\label{Reduct10dRicci}

In this appendix we provide the Ricci tensor and the Ricci scalar of the 10-dimensional metric (\ref{10dmetric}), as well as the reduction of the self-duality relation of the RR field-strength $F_5$. 

The 10-dimensional vielbeine $\{E^\alpha,\,E^{4+a}\}$ are\begin{eqnarray}
\nonumber E^\alpha &=& e^\varphi \rho^\alpha\,, \quad\qquad\qquad\qquad\qquad \alpha = 0,1,\ldots ,4\,,\\ [2mm]
E^{4+a} &=& V^a{}_b\, e^b + \delta^a_5\, e^v A\,, \qquad\qquad a,b = 1,2\ldots ,5\,.
\end{eqnarray}
Here, $\rho^\alpha(x)$ are vielbeine for the 5-dimensional spacetime metric $g_{\mu\nu}$, while
the Weyl rescaling factor $\varphi(x)$, setting the reduced action in the Einstein frame, is
\begin{equation}\label{eq:WeylFactor}
\varphi \,=\, -\frac{1}{6}\,\log(g/g_{\rm SE})\,=\, -\frac{1}{3}(4u+v)    \,,
\end{equation}
where $g :=\det g_{ab}$, and the constant $g_{\rm SE}$ is $g$ evaluated at the Sasaki--Einstein point (\ref{eq:SEpoint}). Moreover, $e^b$ are the coset 1-forms, and the matrix $V^a{}_b(x)$ is such that $V_a{}^c \,\delta_{cd}\,V^d{}_b = g_{ab}\,$; namely, the 1-forms $V^a{}_b e^b$ are vielbeine on $T^{1,1}$. In particular we have $E^9 = e^{v}(\eta + A)$.

We find that in flat indices the 10-dimensional Ricci tensor decomposes as
\begin{eqnarray}
 R^{(10)}_{\alpha\beta} \!\!\!&=&\!\! e^{-2\varphi}\,\Big[R_{\alpha\beta} +\frac{1}{2}\,e^{-2\varphi+2v}F_\alpha{}^\gamma F_{\gamma\beta}-3\partial_\alpha\varphi\partial_\beta\varphi-\eta_{\alpha\beta}\,\square\varphi-\frac{1}{4}g^{ac}g^{bd}D_\alpha g_{ab}\,D_\beta g_{cd}\Big],\nonumber\\[2mm]
R^{(10)}_{\alpha b}\!\!\!&=&\!\! -\frac{1}{2}\,\delta_b^{5}\,e^{-\varphi-v}\,\left[\nabla^\beta\left(e^{-2\varphi+2v}F_{\beta\alpha}\right)-3 \,D_\alpha g_{cd}\,\,\omega^{cd}{}_5\right],\\[2mm]
R^{(10)}_{ab} \!\!\!&=&\!\!\!\! V^{-1\,c}{}_{a} V^{-1\,d}{}_{b}\Big[(R_{T^{1,1}})_{cd} +\frac{1}{2}e^{-2\varphi}\!\left(g^{ef}\,D_\gamma g_{ce} D^\gamma g_{df}-D_\gamma D^\gamma g_{cd}\right)\!\Big]\!+\frac{1}{4}\delta_a^5\delta_b^5e^{-4\varphi+2v}F_{\alpha\beta}F^{\alpha\beta}. \nonumber
 \end{eqnarray}
The flat indices on the left hand side are defined with respect to the 10d vielbeine $\{E^\alpha,\,E^{4+a}\}$, while the indices on the right hand side refer to the frame defined by the 5-dimensional vielbeine $\rho^\alpha$ and the coset 1-forms $e^a$.
Furthermore, we have introduced the field strength $F=dA$, as well as the gauge covariant derivative
\begin{equation}
D g_{ab} \,=\,  dg_{ab} +6 A\,\,\omega_{(ab)5}\,, 
\end{equation}
$\omega^a{}_b \equiv \omega_c{}^a{}_b\,e^c$ being the spin connection on $T^{1,1}$. From the explicit form of $\omega^a{}_b\,$, we see that the connection term modifies only the derivative of $\theta$ in $g_{ab}$, yielding $D\theta = d\theta -3 A$.

Notice that, as a consequence of our left-invariant ansatz, the dependence on the internal coordinates dropped out; this is essential for the consistency of the truncation.

The Ricci scalar reads
\begin{equation}
R^{(10)} \!= e^{-2\varphi}R_M+R_{T^{1,1}} - \frac{1}{4}e^{-4\varphi+2v}F_{\alpha\beta}F^{\alpha\beta} - e^{-2\varphi}\Big[3\partial_\alpha\varphi\partial^\alpha\varphi + 2\square_5\varphi+\frac{1}{4}g^{ac}g^{bd}D_\alpha g_{ab}\,D^\alpha g_{cd}\Big],
\end{equation}
with the corresponding Einstein--Hilbert term reducing to a 5-dimensional action as 
\begin{eqnarray}
\frac{1}{2\kappa_{10}^2}\int (R*1)_{10}\!&=&\!\frac{1}{2\kappa_{5}^2}\int\,\Big[R_M+e^{2\varphi} R_{T^{1,1}}-\frac{1}{4}e^{-2\varphi+2v}\,F_{\alpha\beta}\,F^{\alpha\beta}\nonumber\\ [2mm]
&&\quad\quad -\frac{1}{4}\,g^{ac}\,g^{bd}\, D_\alpha g_{ab} D^\alpha g_{cd} -\frac{1}{12}\,\partial_\alpha\left(\log{g}\right)\partial^\alpha\left(\log{g}\right) \Big]\!*\!1\,,
\end{eqnarray}
where (\ref{eq:WeylFactor}) has been used. The 5-dimensional gravitational coupling is
\begin{equation}\label{eq:5dcoupling}
\kappa_5^2 \,=\, \frac{\kappa_{10}^2}{V_{\rm SE}}\;,\qquad \qquad\quad V_{\rm SE}=\int_{T^{1,1}}\sqrt{g_{\rm SE}}\,e^{12345}=\frac{1}{2}\int_{T^{1,1}} J\wedge J\wedge\eta \,,
\end{equation}
where the reference volume $V_{\rm SE}$ is the coset volume at the Sasaki--Einstein point. By expressing $g_{ab}$ in terms of its elements as in (\ref{GenericLeftInvMetric}), (\ref{eq:ReparamMetric}), we obtain \hbox{eq.~(\ref{S_PureMetric0}) of the main text.}

\vskip 3mm

Another computation involving the metric (\ref{10dmetric}) is the reduction of the self-duality relation $F_5= *F_5$ of the RR 5-form. Recalling (\ref{eq:T11HodgeStar}) for the Hodge star on $T^{1,1}$, this translates in the following 5-dimensional relations:
\begin{eqnarray}
f_5  \!\!\!&=& \!\!-2\,e^{-\frac{32}{3}u-\frac{8}{3}v}*f_0\nonumber\\[2mm]
f_4\!\!\! &=&\! 2\,e^{-8u}*f_1\nonumber\\[2mm]
f_3^J \!\!\!&=&\!\!\!  -e^{-\frac{4}{3}u-\frac{4}{3}v}*\!\Big[\left(\chch t\:\ch{(4w)}- \shsh t\right)f_2^J-\chch t\:\sh(4w)f_2^\Phi - \,\sh(2t)\:\sh{(2w)}\,{\rm Re}(ie^{-i\theta}f_2^\Omega)\Big]\nonumber \\[2mm]
f_3^\Phi \!\!\!&=&\!\!\!  -e^{-\frac{4}{3}u-\frac{4}{3}v}*\!\Big[\chch t\:\sh{(4w)}f_2^J-\left(\chch t\:\ch{(4w)}+ \shsh t\right)f_2^\Phi  - \,\sh(2t)\:\ch{(2w)}\,{\rm Re}(ie^{-i\theta}f_2^\Omega)\Big]\nonumber\\[2mm]
f_3^\Omega \!\!\!&=&\!\!\! -e^{-\frac{4}{3}u-\frac{4}{3}v}*\!\Big[i\,e^{i\theta}\sh(2t)\:\sh{(2w)}f_2^J-i\,e^{i\theta}\sh(2t)\:\ch{(2w)}f_2^\Phi +\, \chch t\,f_2^\Omega- \shsh t\,e^{2i\theta}\,\overline{f_2^\Omega}\Big]. \nonumber\\ [2mm] \label{5dselfduality} 
\end{eqnarray}

\section{Recovering the Papadopoulos--Tseytlin ansatz}\label{RecoveringPT}

The Papadopoulos--Tseytlin (PT) ansatz \cite{PapadopoulosTseytlin} and its 5-dimensional generalization \cite{BergHaackMueck} are naturally incorporated in our supersymmetric truncation. We provide here a dictionary between the fields in that truncation and the corresponding subset of the ones appearing in this work, giving an interpretation to the additional constraint they have. 

Given an explicit parameterization of $T^{1,1}$ in terms of angular coordinates $\{\theta_1,\phi_1,\theta_2,\phi_2,\psi\}$, with ranges $0\leq \theta_{1,2} < \pi$, $\;0\leq \phi_{1,2} < 2\pi$, and $\;0\leq \psi < 4\pi$, we can choose the coframe 1-forms as in \cite{KlebanovStrassler}, namely
\begin{eqnarray}
\nonumber e^1 \!\!&=&\!\! -\sin\theta_1 d\phi_1\,,\qquad e^2 \,=\, d\theta_1\,, \\ [2mm]
\nonumber e^3 \!\!&=&\!\! \cos \psi \sin\theta_2 d\phi_2 - \sin\psi d\theta_2\,,\\ [2mm]
\nonumber e^4 \!\!&=&\!\! \sin\psi\sin\theta_2 d\phi_2 + \cos\psi d\theta_2\,,\\ [2mm]
\label{vielbeins} e^5 \!\!&=&\!\! d\psi +\cos\theta_1 d\phi_1 + \cos\theta_2 d\phi_2\,.
\end{eqnarray}
These are precisely the ones used in \cite{BergHaackMueck} for expanding both the forms and the internal metric. Notice that they satisfy the differential relations (\ref{eq:ExtDerCoframe}) and therefore can be used to define a basis of forms $\{\eta, J, \Omega, \Phi\}$ exactly with the same combinations as (\ref{leftinvforms}). This facilitates the translation of the different quantities to our notation.

In this manner, the set of scalar fields $\{p, x, g, a, b, h_1, h_2, K, \chi\}$ and fluxes $\{P, Q\}$ considered in \cite{PapadopoulosTseytlin, BergHaackMueck} is identified with a subsector of the ones used in this work. In detail, after switching off the flux $p$, all our vector and form fields, as well as the scalars $\{{\rm Re}\,b^\Omega, c^J, c^\Phi, {\rm Im}\,c^\Omega, a, \theta, C_0\}$, the remaining scalars and fluxes are related to the PT ones as follows
\begin{equation}
\begin{array}{c|ccccccccccc}
{\rm PT } & 54e^{-6p} & 6e^x & e^{-g} & a & -6Pb & -6h_1 & -6h_2 & -6\chi & 54K & 18P & 54Q\\ \hline
{\rm here} & e^{2u+2v} & e^{2u} & e^{-2w}\ch\,t & - e^{2w}\tanh{t} & {\rm Re}\,c^\Omega & b^\Phi & {\rm Im}\,b^\Omega & b^J & f_0 & q & k
\end{array} 
\end{equation}
The Hamiltonian constraint imposed in \cite{PapadopoulosTseytlin, BergHaackMueck} (cf. eq. (3.11) of \cite{BergHaackMueck}) to ensure the consistency of the truncation has in our model a natural interpretation: it comes from the equation of motion for $b_1$ (descending from the $B$-field equation of type IIB supergravity), that in our case is
\begin{eqnarray}
d\left(e^{\frac{8}{3}u-\frac{4}{3}v-\phi}\,*\,h_2\right)&=&4e^{-4u-\phi}\,*\Big[\left(\chch t\,\ch(4w)-\shsh t\right)\,h_1^J-\chch t\,\sh(4w)\,h_1^\Phi\nonumber\\[2mm]
&&-\sh(2t)\,\sh(2w)\,{\rm Re}\left(ie^{-i\theta} h_1^\Omega\right)\Big]+e^{\frac{16}{3}u+\frac{4}{3}v-\phi}\,dA\wedge*h_3\nonumber\\[2mm]
&&+e^{\frac{8}{3}u-\frac{4}{3}v+\phi}\,dC_0\wedge*g_2-2g_3\wedge f_1-2g_1^J\wedge f_3^J+2g_1^\Phi\wedge f_3^\Phi\nonumber\\[2mm]
&&-2\,{\rm Re}\left(g_1^\Omega\wedge \overline{f_3^\Omega}\right).
\end{eqnarray}
Switching off the fields not appearing in the PT ansatz, the equation above reduces to the first-order equation
\begin{equation}
db^J=\frac{\chch t\,\sh(4w)\,db^\Phi  -  \sh(2t)\,\sh(2w)\,d\,{\rm Im} \,b^\Omega }{\left(\chch t\,\ch(4w)-\shsh t\right)}\,,
\end{equation}
relating the derivatives of $b^J$, $b^\Phi$ and ${\rm Im}\,b^\Omega$. Once the dictionary is used, it reproduces the constraint in PT. The rest of the equations of motion for the fields not present in the PT ansatz are exactly satisfied and thus do not give supplementary constraints.

\section{The gauge transformations}\label{GaugeTransf}

In this appendix we study how the gauge symmetry of the 5-dimensional action arises from the 10-dimensional reparameterization invariance as well as from the gauge symmetry of the type IIB supergravity form fields.

As already discussed in section \ref{Red10dCurvature}, a crucial symmetry of the action comes from the invariance under reparametrizations of the U(1) fibre coordinate which are local in the \hbox{5-dimensional} spacetime, $\psi\to\psi -3\,\omega(x)$. Then the associated 1-form $\eta \equiv -\frac{1}{3}e^5\,$ is shifted as $\,\delta \eta = d\omega\,$.
Recalling the ansatz for the 10-dimensional metric (\ref{10dmetric}), this is interpreted from the \hbox{5-dimensional} viewpoint as a gauge transformation of the 1-form $A$,
\begin{equation}
\delta A\,=\,d\omega\,.
\end{equation}
We already saw that among the metric parameters we have a charged scalar, whose phase $\theta$ shifts as
\begin{equation}
\delta \theta\, =\, 3\,\omega\,.
\end{equation}
Furthermore, one can see from the explicit expression of the coset 1-forms (\ref{vielbeins}) that while $J$ and $\Phi$, defined in (\ref{leftinvforms}), are invariant, $\Omega$ has an explicit dependence on the coordinate $\psi$, and thus transforms. This implies that the spacetime fields associated to $\Omega$ in the expansion of the type IIB forms are charged. Applying this to the 3-form field strengths $H$ and $F_3$ expanded as in section \ref{FormFields}, one deduces the following transformations
\begin{equation}\label{eq:ScalTransfs1}
\begin{array}{rclcrcl}
\delta b^\Phi & = & p\,\omega\,, & \quad\quad\quad\quad & \delta c^\Phi & = & q\,\omega\,,\\[2mm]
\delta b^\Omega & = & 3i\omega\,b^\Omega\,, & \quad\quad\quad\quad & \delta c^\Omega & = & 3i\omega\,c^\Omega\,.
\end{array}
\end{equation}
We point out that, while $b^\Omega$ and $c^\Omega$ have the charges and couplings of abelian Higgs scalars, the real fields $b^\Phi$ and $c^\Phi$ have shift transformations and are therefore St\"{u}ckelberg coupled, with the corresponding covariant derivatives $Db^\Phi,D c^{\Phi}$(given in section \ref{FormFields}) being invariant.

We now pass to consider the gauge symmetries of the type IIB forms. Together with the usual transformations for the 2-forms
\begin{equation}
\delta B=d\Lambda\,,\quad\quad\quad\delta C_2=d\Gamma
\end{equation}
we see that the presence of the 3-form fluxes alters the transformation of the RR 4-form. In particular, imposing invariance of the five-form $F_5$ we obtain the transformation for the derivative
\begin{equation}
\delta dC_4=\frac{1}{2}\left[d\Gamma\wedge\left(dB+2H^{\rm fl}\right)-d\Lambda\wedge\left(dC_2+2F_3^{\rm fl}\right)\right]
\end{equation} 
that we can solve as
\begin{equation}
\delta C_4=d\Sigma+\frac{1}{2}\left[d\Gamma\wedge B-d\Lambda\wedge C_2\right]+\Gamma\wedge H^{\rm fl}-\Lambda\wedge F_3^{\rm fl},
\end{equation}
where $\Sigma$ is a 3-form on the 10-dimensional spacetime. Expanding the 10-dimensional gauge parameters in our basis of forms,
\begin{eqnarray}
\nonumber \Lambda &=&\lambda_1+\lambda\,\left(\eta+A\right),\quad\qquad \Gamma \;=\;\gamma_1+\gamma\,\left(\eta+A\right),\\ [2mm]
\Sigma &=& (\sigma^J \wedge J + \sigma^\Phi \wedge \Phi)\wedge (\eta +A) + {\rm Re}\big[ \sigma^\Omega\wedge \Omega\wedge (\eta+A) + \sigma_1^\Omega\wedge \Omega \big] + \ldots\,,
\end{eqnarray}
with 5-dimensional gauge parameters $\lambda,\,\sigma,\,\gamma$, we complete the set of transformations of the charged scalars
\begin{eqnarray}
\delta b^J \!& = &\! 2\lambda\,, \qquad\qquad\quad \delta c^J  \,=\,  2\gamma\,, \nonumber\\[1mm]
 \delta a \!& = &\! 2\sigma^J+\gamma\, b^J-\lambda\, c^J+k\,\omega + \frac{1}{2}\omega\,(p\,c^\Phi-q \,b^\Phi)\,, \label{eq:ScalTransfs2}\end{eqnarray}
as well as of the 1-forms
\begin{eqnarray}
 \delta b_1 & = & d\lambda\,,\hspace{4.5 cm}\;\, \delta c_1  =  d\gamma\,,\nonumber\\[1mm]
 \delta \tilde{a}^J_1 & = & d\sigma^J-2\lambda\, c_1+2\gamma\, b_1\,, \qquad\qquad  \delta \tilde{a}_1^\Omega  = D\sigma^\Omega - 3i \sigma_1^\Omega + 3i\omega\,\tilde{a}_1^\Omega\nonumber\\[1mm]
 \delta \tilde{a}_1^\Phi & = & d\sigma^\Phi+p\left(\gamma_1+\gamma\,A-\omega\,c_1\right)-q\left(\lambda_1+\lambda\,A-\omega\,b_1\right)\,,\quad
\end{eqnarray}
and the 2-forms
\begin{equation}
 \delta b_2  =  d\lambda_1+\lambda\,dA\,,\qquad\quad
 \delta c_2  =  d\gamma_1+\gamma\, dA,\qquad\quad
 \delta \tilde{a}_2^\Omega  =  D\sigma_1^\Omega + \sigma^\Omega dA +  3i\omega\,\tilde{a}_2^\Omega\,.
\end{equation}
Here, we implemented the field redefinitions
\begin{equation}\label{eq:FieldRedef}
\tilde{a}_1^{\{J,\Phi,\Omega\}}=a_1^{\{J,\Phi,\Omega\}}-\frac{1}{2}b^{\{J,\Phi,\Omega\}}\,c_1+\frac{1}{2}c^{\{J,\Phi,\Omega \}}\,b_1\,,\quad\quad\quad\tilde{a}_2^\Omega=a_2^\Omega-\frac{1}{2}b^\Omega c_2+\frac{1}{2}c^\Omega b_2\,,
\end{equation}
which remove some scalar-dependent terms present in the variations of the old fields. These redefined fields are the ones directly identified with the $\mathcal{N}=4$ vectors in the body of the paper. As a further check of the validity of the gauged $\mathcal{N}=4$ supergravity picture we have matched these variations with the ones provided in \cite{SchonWeidner} and given essentially in terms of the embedding tensor (see also \cite{IIBonSE} for a discussion from this perspective).

One can check that the 5-dimensional field strengths $h_p$, $g_p$ and $f_p$ defined in the main text are all covariant under these form transformations.

The gauge group $G$ of our 5-dimensional model can be deduced by studying the scalar transformations. From (\ref{eq:ScalTransfs1}), (\ref{eq:ScalTransfs2}) we see that the only non-vanishing commutator is
\begin{equation}
[\delta_\lambda,\delta_\gamma]\,a\, =4\lambda\gamma\,=\, 2\,\delta_{(\sigma^J=\lambda\gamma)}\,a\,,
\end{equation}
hence we conclude that $G={\rm Heis}_3\times {\rm U(1)}$, with the Heisenberg factor being generated by the transformations with parameters $\lambda, \gamma, \sigma^J$, and the U(1) being generated by the \hbox{$\omega$-transformations}. This agrees with what found in section  \ref{EmbTensor} via the embedding tensor. The realization of the gauge group on the 1-forms is more subtle, due to the entanglement with the 2-forms: while $\{A, b_1, c_1, \tilde{a}_1^J\}$ are proper gauge fields in the adjoint representation of $G$, the variations of $\tilde a_1^\Omega$ and $\tilde a_1^\Phi$ also contain the 2-form transformation parameters, and therefore do not correspond to standard Lie algebra transformations.

Finally, we remark that by switching off the NSNS and RR fluxes, as well as the geometric fluxes associated with the non-closure of the basis forms,  we obtain the correct limit to ungauged $\mathcal N=4$ supergravity. Indeed, setting $k=p=q=0$, and repeating the derivation above assuming $d\eta=d\Omega=0$, the gauge variations of the 5-dimensional forms reduce to abelian transformations, and the scalars become all neutral.



\begin{thebibliography}{80}

\bibitem{KlebanovWitten}
  I.~R.~Klebanov and E.~Witten,
  {\it Superconformal field theory on threebranes at a Calabi-Yau  singularity},
  Nucl.\ Phys.\  B {\bf 536}, 199 (1998)
  [arXiv:hep-th/9807080].

\bibitem{CandelasOssaConifolds}
  P.~Candelas and X.~C.~de la Ossa,
  {\it Comments on Conifolds},
  Nucl.\ Phys.\  B {\bf 342} (1990) 246.

\bibitem{KlebanovNekrasov}
  I.~R.~Klebanov and N.~A.~Nekrasov,
  {\it Gravity duals of fractional branes and logarithmic RG flow},
  Nucl.\ Phys.\  B {\bf 574} (2000) 263
  [arXiv:hep-th/9911096].

\bibitem{KlebanovTseytlin}
  I.~R.~Klebanov and A.~A.~Tseytlin,
  {\it Gravity Duals of Supersymmetric SU(N)$\times$SU(N+M) Gauge Theories},
  Nucl.\ Phys.\  B {\bf 578} (2000) 123
  [arXiv:hep-th/0002159].

\bibitem{PandoZayasTseytlin}
  L.~A.~Pando Zayas and A.~A.~Tseytlin,
  {\it 3-branes on resolved conifold},
  JHEP {\bf 0011} (2000) 028
  [arXiv:hep-th/0010088].

\bibitem{KlebanovStrassler}
  I.~R.~Klebanov and M.~J.~Strassler,
  {\it Supergravity and a confining gauge theory: Duality cascades and
  chiSB-resolution of naked singularities},
  JHEP {\bf 0008} (2000) 052
  [arXiv:hep-th/0007191].
  
\bibitem{MaldacenaNunez}
  J.~M.~Maldacena and C.~Nunez,
  {\it Towards the large N limit of pure N = 1 super Yang Mills},
  Phys.\ Rev.\ Lett.\  {\bf 86}, 588 (2001)
  [arXiv:hep-th/0008001].
  
\bibitem{MaldacenaMartelli}
  J.~Maldacena, D.~Martelli,
  {\it The Unwarped, resolved, deformed conifold: Fivebranes and the baryonic branch of the Klebanov-Strassler theory},
  JHEP {\bf 1001 } (2010)  104.
  [arXiv:0906.0591 [hep-th]].

\bibitem{GPPZ}
  L.~Girardello, M.~Petrini, M.~Porrati and A.~Zaffaroni,
  {\it Novel local CFT and exact results on perturbations of N = 4 super
  Yang-Mills from AdS dynamics},
  JHEP {\bf 9812} (1998) 022
  [arXiv:hep-th/9810126].
  
\bibitem{FGPW}
  D.~Z.~Freedman, S.~S.~Gubser, K.~Pilch and N.~P.~Warner,
  {\it Renormalization group flows from holography supersymmetry and a
  c-theorem},
  Adv.\ Theor.\ Math.\ Phys.\  {\bf 3} (1999) 363
  [arXiv:hep-th/9904017].

\bibitem{GPPZ2} L.~Girardello, M.~Petrini, M.~Porrati and A.~Zaffaroni,
  {\it The supergravity dual of N = 1 super Yang-Mills theory},
  Nucl.\ Phys.\  B {\bf 569} (2000) 451
  [arXiv:hep-th/9909047].

\bibitem{PapadopoulosTseytlin}
  G.~Papadopoulos, A.~A.~Tseytlin,
  {\it Complex geometry of conifolds and five-brane wrapped on two sphere},
  Class.\ Quant.\ Grav.\  {\bf 18 } (2001)  1333-1354.
  [hep-th/0012034].
  
\bibitem{BergHaackMueck}
  M.~Berg, M.~Haack, W.~Mueck,
  {\it Bulk dynamics in confining gauge theories},
  Nucl.\ Phys.\  {\bf B736 } (2006)  82-132.
  [hep-th/0507285].

\bibitem{BaryonicBranchKS}
  A.~Butti, M.~Grana, R.~Minasian, M.~Petrini and A.~Zaffaroni,
  {\it The baryonic branch of Klebanov-Strassler solution: A supersymmetric
  family of SU(3) structure backgrounds},
  JHEP {\bf 0503}, 069 (2005)
  [arXiv:hep-th/0412187].
  
\bibitem{FakeSugra}
  D.~Z.~Freedman, C.~Nunez, M.~Schnabl and K.~Skenderis,
  {\it Fake Supergravity and Domain Wall Stability},
  Phys.\ Rev.\  D {\bf 69} (2004) 104027
  [arXiv:hep-th/0312055].
  
\bibitem{IIBonSE}
  D.~Cassani, G.~Dall'Agata and A.~F.~Faedo,
  {\it Type IIB supergravity on squashed Sasaki-Einstein manifolds},
  JHEP {\bf 1005} (2010) 094
  [arXiv:1003.4283 [hep-th]].

\bibitem{LiuEtAl}
  J.~T.~Liu, P.~Szepietowski and Z.~Zhao,
  {\it Consistent massive truncations of IIB supergravity on Sasaki-Einstein
  manifolds},
  Phys.\ Rev.\  D {\bf 81}, 124028 (2010)
  [arXiv:1003.5374 [hep-th]].

\bibitem{GauntlettVarelaSE}
  J.~P.~Gauntlett and O.~Varela,
  {\it Universal Kaluza-Klein reductions of type IIB to N=4 supergravity in five
  dimensions},
  JHEP {\bf 1006}, 081 (2010)
  [arXiv:1003.5642 [hep-th]].
  
\bibitem{SkenderisTaylorTsimpis}
  K.~Skenderis, M.~Taylor and D.~Tsimpis,
  {\it A consistent truncation of IIB supergravity on manifolds admitting a
  Sasaki-Einstein structure},
  JHEP {\bf 1006} (2010) 025
  [arXiv:1003.5657 [hep-th]].
  
\bibitem{BuchelLiu}
  A.~Buchel and J.~T.~Liu,
  {\it Gauged supergravity from type IIB string theory on Y(p,q)
  manifolds},  Nucl.\ Phys.\  B {\bf 771} (2007) 93  [arXiv:hep-th/0608002].

\bibitem{GauntlettVarela07}
  J.~P.~Gauntlett and O.~Varela,
  {\it Consistent Kaluza-Klein Reductions for General Supersymmetric
  AdS Solutions},  Phys.\ Rev.\  D {\bf 76} (2007) 126007
  [arXiv:0707.2315 [hep-th]].  
  
\bibitem{MaldacenaMartelliTachikawa}
  J.~Maldacena, D.~Martelli and Y.~Tachikawa,
  {\it Comments on string theory backgrounds with non-relativistic conformal
  symmetry},
  JHEP {\bf 0810} (2008) 072
  [arXiv:0807.1100 [hep-th]].
  
\bibitem{GauntlettKimVarelaWaldram}  
  J.~P.~Gauntlett, S.~Kim, O.~Varela and D.~Waldram,  
  {\it Consistent supersymmetric Kaluza--Klein truncations with massive modes},
  JHEP {\bf 0904} (2009) 102
  [arXiv:0901.0676 [hep-th]].

\bibitem{Gubser1}
  S.~S.~Gubser, C.~P.~Herzog, S.~S.~Pufu and T.~Tesileanu,
  {\it Superconductors from Superstrings},
  Phys.\ Rev.\ Lett.\  {\bf 103} (2009) 141601
  [arXiv:0907.3510 [hep-th]].

\bibitem{ExploitingN=2}
  D.~Cassani and A.~K.~Kashani-Poor,
  {\it Exploiting N=2 in consistent coset reductions of type IIA},
  Nucl.\ Phys.\  B {\bf 817} (2009) 25
  [arXiv:0901.4251 [hep-th]].

\bibitem{T11spectrum}
  A.~Ceresole, G.~Dall'Agata, R.~D'Auria and S.~Ferrara,
  {\it Spectrum of type IIB supergravity on AdS(5)$\,\times\,$T(11): Predictions on N  = 1
  SCFT's},
  Phys.\ Rev.\  D {\bf 61} (2000) 066001
  [arXiv:hep-th/9905226].

\bibitem{RomansNewIIBsol}
  L.~J.~Romans,
  {\it New Compactifications Of Chiral N=2 D = 10 Supergravity},
  Phys.\ Lett.\  B {\bf 153} (1985) 392.

\bibitem{Ramallo}
 F.~Benini, F.~Canoura, S.~Cremonesi, C.~Nunez and A.~V.~Ramallo,
 {\it Backreacting Flavors in the Klebanov-Strassler Background},
 JHEP {\bf 0709} (2007) 109
 [arXiv:0706.1238 [hep-th]].

\bibitem{HerzKlebPufuTesi}
  C.~P.~Herzog, I.~R.~Klebanov, S.~S.~Pufu and T.~Tesileanu,
  {\it Emergent Quantum Near-Criticality from Baryonic Black Branes},
  JHEP {\bf 1003} (2010) 093
  [arXiv:0911.0400 [hep-th]].

\bibitem{Bena:2010pr}
  I.~Bena, G.~Giecold, M.~Grana, N.~Halmagyi and F.~Orsi,
  {\it Supersymmetric Consistent Truncations of IIB on T(1,1)},
  arXiv:1008.0983 [hep-th].

\bibitem{MuellerStuckl}
  F.~Mueller-Hoissen and R.~Stuckl,
  {\it Coset spaces and ten-dimensional unified theories},
  Class.\ Quant.\ Grav.\  {\bf 5} (1988) 27.

\bibitem{CastDAuriaFreBook}
  L.~Castellani, R.~D'Auria and P.~Fre,
  {\it Supergravity And Superstrings: A Geometric Perspective. Vol. 1:
  Mathematical Foundations},
Singapore, World Scientific (1991).

\bibitem{MinasianTsimpis}
  R.~Minasian and D.~Tsimpis,
  {\it On the geometry of non-trivially embedded branes},
  Nucl.\ Phys.\  B {\bf 572} (2000) 499
  [arXiv:hep-th/9911042].
  
\bibitem{KachruKashaniPoor}
S.~Kachru and A.~K.~Kashani-Poor,
  {\it Moduli potentials in type IIA compactifications with RR and NS flux},
  JHEP {\bf 0503} (2005) 066
  [arXiv:hep-th/0411279].

\bibitem{IIAModuliStabilization}
O.~DeWolfe, A.~Giryavets, S.~Kachru and W.~Taylor,
  {\it Type IIA moduli stabilization},
  JHEP {\bf 0507}, 066 (2005)
  [arXiv:hep-th/0505160].

\bibitem{ReducingSU3SU3} 
  D.~Cassani,
  {\it Reducing democratic type II supergravity on SU(3) $\times$ SU(3) structures},
  JHEP {\bf 0806} (2008) 027
  [arXiv:0804.0595 [hep-th]].

\bibitem{DHZ}
  G.~Dall'Agata, C.~Herrmann and M.~Zagermann,
  {\it General matter coupled N = 4 gauged supergravity in five dimensions},
  Nucl.\ Phys.\  B {\bf 612} (2001) 123
  [arXiv:hep-th/0103106].

\bibitem{SchonWeidner}  J.~Sch\"on and M.~Weidner,  {\it Gauged N = 4 supergravities},
  JHEP {\bf 0605} (2006) 034
  [arXiv:hep-th/0602024].

\bibitem{Dauria}
 R.~D'Auria and P.~Fre,
 {\it Universal Bose-Fermi Mass Relations In Kaluza-Klein Supergravity And
 Harmonic Analysis On Coset Manifolds With Killing Spinors},
 Annals Phys.\  {\bf 162} (1985) 372.

\bibitem{Solvable}
  L.~Andrianopoli, R.~D'Auria, S.~Ferrara, P.~Fre and M.~Trigiante,
  {\it R-R scalars, U-duality and solvable Lie algebras},
  Nucl.\ Phys.\  B {\bf 496} (1997) 617
  [arXiv:hep-th/9611014].

\bibitem{SamtlebenLectures}
  H.~Samtleben,
  {\it Lectures on Gauged Supergravity and Flux Compactifications},
  Class.\ Quant.\ Grav.\  {\bf 25} (2008) 214002
  [arXiv:0808.4076 [hep-th]].

\bibitem{SusyAdSIIB}
  J.~P.~Gauntlett, D.~Martelli, J.~Sparks and D.~Waldram,
  {\it Supersymmetric AdS(5) solutions of type IIB supergravity},
  Class.\ Quant.\ Grav.\  {\bf 23} (2006) 4693
  [arXiv:hep-th/0510125].

\bibitem{PopeWarner}
  C.~N.~Pope and N.~P.~Warner,
  {\it An SU(4) Invariant Compactification Of D = 11 Supergravity On A Stretched
  Seven Sphere},
  Phys.\ Lett.\  B {\bf 150} (1985) 352;
  {\it Two New Classes Of Compactifications Of D = 11 Supergravity},
  Class.\ Quant.\ Grav.\  {\bf 2} (1985) L1.

\bibitem{FreundRubin}
  P.~G.~O.~Freund and M.~A.~Rubin,
  {\it Dynamics Of Dimensional Reduction},
  Phys.\ Lett.\  B {\bf 97} (1980) 233.

\bibitem{Alekseevsky}
D.~V.~Alekseevsky, I.~Dotti and C.~Ferraris, {\it Homogeneous Ricci positive 5-manifolds}, Pacific J. Math. {\bf 175} (1996) 1-12.

\bibitem{BFbound}
  P.~Breitenlohner and D.~Z.~Freedman,
  {\it Positive Energy In Anti-De Sitter Backgrounds And Gauged Extended
  Supergravity},
  Phys.\ Lett.\  B {\bf 115} (1982) 197; 
  {\it Stability In Gauged Extended Supergravity},
  Annals Phys.\  {\bf 144} (1982) 249.
  
\bibitem{MezincescuTownsend}
  L.~Mezincescu and P.~K.~Townsend,
  {\it Stability At A Local Maximum In Higher Dimensional Anti-De Sitter Space And
  Applications To Supergravity},
  Annals Phys.\  {\bf 160} (1985) 406.


\bibitem{BaumannEtAlD3potentials}
  D.~Baumann, A.~Dymarsky, S.~Kachru, I.~R.~Klebanov and L.~McAllister,
  {\it D3-brane Potentials from Fluxes in AdS/CFT},
  JHEP {\bf 1006} (2010) 072
  [arXiv:1001.5028 [hep-th]].
  
\bibitem{CDSWchiralrings}
  F.~Cachazo, M.~R.~Douglas, N.~Seiberg and E.~Witten,
  {\it Chiral Rings and Anomalies in Supersymmetric Gauge Theory},
  JHEP {\bf 0212} (2002) 071
  [arXiv:hep-th/0211170].
  
\bibitem{KlebanovOuyangWitten}
  I.~R.~Klebanov, P.~Ouyang and E.~Witten,
  {\it A gravity dual of the chiral anomaly},
  Phys.\ Rev.\  D {\bf 65} (2002) 105007
  [arXiv:hep-th/0202056].

\bibitem{KlebanovWitten2}
  I.~R.~Klebanov and E.~Witten,
  {\it AdS/CFT correspondence and symmetry breaking},
  Nucl.\ Phys.\  B {\bf 556} (1999) 89
  [arXiv:hep-th/9905104].

\bibitem{CeresoleDallAgataGaugedN=2}
  A.~Ceresole and G.~Dall'Agata,
  {\it General matter coupled N = 2, D = 5 gauged supergravity},
  Nucl.\ Phys.\  B {\bf 585} (2000) 143
  [arXiv:hep-th/0004111].
  
\bibitem{KarchKatz}
  A.~Karch and E.~Katz,
  {\it Adding flavor to AdS/CFT},
  JHEP {\bf 0206} (2002) 043
  [arXiv:hep-th/0205236].

\bibitem{KKreview}
  M.~J.~Duff, B.~E.~W.~Nilsson and C.~N.~Pope,
  {\it Kaluza-Klein Supergravity},
  Phys.\ Rept.\  {\bf 130} (1986) 1.

\bibitem{Klebanov11d}
  I.~R.~Klebanov, S.~S.~Pufu and T.~Tesileanu,
  {\it Membranes with Topological Charge and AdS4/CFT3 Correspondence},
  Phys.\ Rev.\  D {\bf 81} (2010) 125011
  [arXiv:1004.0413 [hep-th]].

\bibitem{Ypq}
  J.~P.~Gauntlett, D.~Martelli, J.~Sparks and D.~Waldram,
  {\it Sasaki-Einstein metrics on $S(2) \times S(3)$},
  Adv.\ Theor.\ Math.\ Phys.\  {\bf 8} (2004) 711
  [arXiv:hep-th/0403002].

\bibitem{Lpqr}
  M.~Cvetic, H.~Lu, D.~N.~Page and C.~N.~Pope,
  {\it New Einstein-Sasaki spaces in five and higher dimensions},
  Phys.\ Rev.\ Lett.\  {\bf 95} (2005) 071101
  [arXiv:hep-th/0504225].



  
\end{thebibliography}
\end{document}